\documentclass[]{aa}  
\usepackage{graphicx}
\usepackage{txfonts}
\usepackage{lipsum}
\usepackage[dvipsnames]{xcolor}
\usepackage{stfloats}

\usepackage{array}
\newcolumntype{L}[1]{>{\raggedright\let\newline\\\arraybackslash\hspace{0pt}}m{#1}}
\newcolumntype{C}[1]{>{\centering\let\newline\\\arraybackslash\hspace{0pt}}m{#1}}
\newcolumntype{R}[1]{>{\raggedleft\let\newline\\\arraybackslash\hspace{0pt}}m{#1}}
\newcommand\Tstrut{\rule{0pt}{2.6ex}}         % = `top' strut
\newcommand\Bstrut{\rule[-1.5ex]{0pt}{0pt}}   % = `bottom' strut
\newcommand*{\dprime}{^{\prime\prime}\mkern-1.2mu}

\newcommand{\cott}{$^{13}$CO }
\newcommand{\coet}{C$^{18}$O }

\begin{document} 
   \title{Tracing pebble drift and trapping using radial carbon depletion profiles in protoplanetary disks}

   \author{J.A. Sturm
          \inst{1}
          \and
          M.K. McClure\inst{1}
          \and
          D. Harsono
          \inst{2}
          \and
          S. Facchini
          \inst{3,4}
          \and
          F. Long
          \inst{5}
          \and
          M. Kama
          \inst{6,7}
          \and
          E.A. Bergin
          \inst{8}
          \and
          E.F. van Dishoeck
          \inst{1,9}
          }

   \institute{
            Leiden Observatory, Leiden University, P.O. Box 9513, NL-2300 RA Leiden, The Netherlands, e-mail: sturm@strw.leidenuniv.nl
            \and
            Institute of Astronomy and Astrophysics, Academia Sinica, No. 1, Sec. 4, Roosevelt Road, Taipei 10617, Taiwan, R. O. C.
            \and
            European Southern Observatory, Karl-Schwarzschild-Strasse 2, 85748 Garching bei M\"unchen, Germany
            \and
            Dipartimento di Fisica, Universit\`a degli Studi di Milano, via Celoria 16, 20133 Milano, Italy
            \and
            Harvard-Smithsonian Center for Astrophysics, 60 Garden Street, Cambridge, MA 02138, USA
            \and
            Department of Physics and Astronomy, University College London, Gower Street, London, WC1E 6BT, UK
            \and
            Tartu Observatory, Observatooriumi 1, T\~oravere 61602, Tartu, Estonia
            \and
            Department of Astronomy, The University of Michigan, 500 Church St., 830 Dennison Bldg., Ann Arbor, MI 48109, USA
            \and
            Max-Planck-Institut f{\"u}r extraterrestrische Physik, Giessenbachstra{\ss}e 1, 85748 Garching bei M\"unchen, Germany
            }
    \date{Received XXX; accepted YYY}

    \abstract
   {The composition of planets may be largely determined by the chemical processing and accretion of icy pebbles in protoplanetary disks. 
   Recent observations of protoplanetary disks hint at wide-spread depletion of gaseous carbon. 
   The missing volatile carbon is likely frozen in CO and/or CO$_2$ ice on grains and locked into the disk through pebble trapping in pressure bumps or planetesimals.}
   {We aim to measure the total elemental C/H ratio in the outer region of seven disks, four of which have been previously shown to be depleted of carbon gas interior to 0.1 AU, through near-infrared spectroscopy.}
   {We present the results of the first successful ACA (Atacama Compact Array) [C~I] $J$ = 1-0 mini-survey of seven protoplanetary disks.
   Using tailored azimuthally symmetric DALI (Dust And LInes) thermo-chemical disk models, supported by the [C I] $J$ = 1-0 and resolved CO isotopologue data, we determine the system-averaged elemental volatile carbon abundance in the outer disk of three sources.}
   {Six out of seven sources are detected in [C~I] $J$ = 1-0 with ACA, four of which show a distinct disk component. 
   Based on the modeling we find severe cold gaseous carbon depletion by a factor of 157$^{+17}_{-15}$ in the outer disk of DL Tau and moderate depletion in the outer disks of DR Tau and DO Tau, by factors of 5$^{+2}_{-1}$ and 17$^{+3}_{-2}$, respectively.
   The carbon abundance is in general expected to be higher in the inner disk if carbon-rich ices drift on large grains towards the star. 
   Combining the outer and inner disk carbon abundances, we demonstrate definitive evidence for radial drift in the disk of DL Tau, where the existence of multiple dust rings points to either short lived or leaky dust traps. We find dust locking in the compact and smooth disks of DO Tau and DR Tau, hinting at unresolved dust substructure. Comparing our results with inner/outer disk carbon depletion around stars of different ages and luminosities, we identify an observational evolutionary trend in gaseous carbon depletion that is consistent with dynamical models of CO depletion processes.}
   {Transport efficiency of solids in protoplanetary disks can significantly differ from what we expect based on the current resolved substructure in the continuum observations. This has important implications for our understanding of the impact of radial drift and pebble accretion on planetary compositions.}

    \maketitle
%-------------------------------------------------------------------
%\vspace*{-1 mm}\
\section{Introduction}
\label{sec:intro}
Carbon and oxygen are two of the most abundant elements in interstellar clouds and play a crucial role in the chemistry of star forming regions and planetary atmospheres \citep{2009ASPC..417...71L,2019ARA&A..57..617M}.
If volatile carbon is carried mostly by dust grains, these grains can grow to produce planetesimals rich in complex organic molecule ices, the building blocks of life.
CO isotopologue surveys have recently revealed that CO is frequently underabundant by orders of magnitude in the gas-phase in protoplanetary disks of T Tauri stars \citep[][]{2001A&A...377..566V,2003A&A...402.1003D,2008A&A...488..565C,2013ApJ...776L..38F,2016ApJ...823...91S,2016ApJ...828...46A,2018ApJ...859...21A,2017ApJ...844...99L,2017A&A...599A.113M}.
The amount of volatile depletion observed cannot be explained solely by freeze-out, photodissociation or chemical processing in the warm molecular layer \citep{2001A&A...377..566V,2014ApJ...788...59W,2017A&A...599A.113M, 2018ApJ...856...85S, 2018A&A...618A.182B, 2019ApJ...883...98Z}. 
Detailed analysis of the TW Hya disk, using [C~I], HD, CO and C$_2$H emission, shows that the depletion in CO is related to elemental carbon depletion \citep{2013Natur.493..644B,2016A&A...592A..83K,2021ApJ...908....8C} and cannot be due to high dust-to-gas ratios only.

\begin{table*}[!b]
    \caption{Host Stellar Properties}
    \label{tab:stellar_props}    
    \begin{tabular}{L{1.9cm}C{1.cm}C{1.0cm}C{1.5cm}C{1cm}C{1cm}C{1cm}C{1cm}C{1cm}C{1.5cm}C{1.5cm}}
    \hline
    \hline
    Name   \Tstrut &SpT\Tstrut&Age\Tstrut& $d$ \Tstrut& $A_\mathrm{V}$\Tstrut & $L_\mathrm{*}$\Tstrut & $T_{\mathrm{eff}}$ \Tstrut& $R_{*}$\Tstrut& $M_{*}$\Tstrut& $\dot{M}$ \Tstrut& $V_{\mathrm{lsr}}$\Tstrut\\
            & &(Myr)&(pc) &(mag)& (L$_\odot$) & (K) & (R$_\odot$) & (M$_\odot$) & (M$_\odot \mathrm{yr}^{-1}$) & (km s$^{-1}$)\\
    (1)  \Bstrut\Tstrut   & (2)\Bstrut\Tstrut & (3) \Bstrut\Tstrut& (4) \Bstrut\Tstrut& (5) \Bstrut\Tstrut& (6) \Bstrut\Tstrut& (7) \Bstrut\Tstrut& (8) \Bstrut\Tstrut& (9) \Bstrut\Tstrut &(10) \Bstrut\Tstrut &(11) \Bstrut\Tstrut   \\
    \hline
    DL Tau\Tstrut  & K7V\Tstrut &7.8\Tstrut & 159 $\pm$ 1    \Tstrut & 1.6 \Tstrut& 0.37 \Tstrut& 4060 \Tstrut& 1.24 \Tstrut& 0.80 \Tstrut& 5.8 x 10$^{-8}$\Tstrut& 6.62 $^{(a)}$\Tstrut\\
    DO Tau  &M0V&2.0& 139 $\pm$ 1     & 3.6 & 0.58 & 3850 & 1.72 & 0.57 & 6.6 x 10$^{-8}$ & 5.91 $^{(b)}$\\
    DR Tau  &M0V&0.9& 195 $^{+3}_{-2}$& 2.1 & 1.12 & 3850 & 2.39 & 0.56 & 4.8 x 10$^{-7}$ & 9.79 $^{(a)}$\\
    FZ Tau  &M0V&1.1& 130 $\pm$ 1     & 6.5 & 0.93 & 3850 & 2.18 & 0.56 & 3.5 x 10$^{-7}$ & 6.0 $^{(c)}$\\
    AS 205 A $^{(d)}$ &K5V&1.1&   128 $\pm$ 2     & 2.9  &2.14  &4266 &2.68&0.87&4.0x10$^{-8}$& 4.583  $^{(e)}$ \\
    FM Cha $^{(f)}$&K7V &5.3&   201 $\pm$ 6     &4.3&0.48&4060&1.40&0.78&8.3x 10$^{-8}$& 4.0  $^{(g)}$\\
    WW Cha  $^{(h)}$\Bstrut&K0V&6.4&   190 $\pm$ 1\Bstrut     &4.1&2.68&5110&2.09&1.65&2.1x 10$^{-7}$&   4.3 $^{(g)}$\Bstrut \\
    \hline
    \end{tabular}
    \\
    \Tstrut
    \textbf{Notes.} Columns are defined as: (1) Name of the central star, (2) stellar spectral type, (3) stellar age, taken from the evolutionary tracks of \citet{1995ApJS..101..117K}, (4) distance to the source based on the Gaia DR2 supplemental catalog \citep{2018AJ....156...58B}, (5) extinction along the line of sight, (6) stellar luminosity, (7) effective stellar temperature, (8) stellar radius, (9) stellar mass, (10) stellar mass accretion rate, (11) local standard of rest velocity of the source used in this work, based on literature CO data.
    \\
    \Tstrut
    \textbf{References.} All stellar quantities of the sources in Taurus are taken from \citet{2019A&A...632A..32M}. (a) \citet{2013A&A...549A..92G}, (b) \citet{2020AJ....159..171F}, (c) taken as the velocity of the maximum [C~I] cloud absorption, (d) \citet{2018ApJ...869L..41A}, (e) \citet{2018ApJ...869L..44K}, (f) \citet{2017A&A...604A.127M}, (g) \citet{2017ApJ...844...99L} and references therein, (h) \citet{2016A&A...585A.136M}
\end{table*}\
The outer disk gas phase carbon depletion can be explained by freeze-out of CO onto dust grains, coupled with a combination of chemical processing into less volatile carbon carrying molecules and dust evolution locking the carbon in the ice on large dust grains in the midplane \citep{2016A&A...592A..83K,2020ApJ...899..134K}.
The long timescales for carbon depletion processes are expected to result in stronger depletion, of several orders of magnitude, in older Class II disks \citep{2020ApJ...899..134K}.
Efficient vertical gas mixing is necessary to replenish the CO gas available for freeze-out near the disk midplane, which results eventually in lower carbon abundances in the atmospheric layers of the disk. 
Large dust grains eventually drift radially inwards towards the central star, due to the gas pressure \citep{2002ApJ...581.1344T,2021A&A...649A..95T}.
If nothing stops the radial drift, the CO ice is released inside the CO snowline resulting in super-solar C/H ratios as observed in HD 163296 \citep{2019ApJ...883...98Z,2020ApJ...891L..16Z}.

However, \citet{2019ApJ...883...98Z} show that not all protoplanetary disks may have an increase in carbon content inside the CO snowline, as a result of carbon locking into less volatile species in the outer disk.
Thus, for example, if CO is processed into CO$_2$ this carbon will not return to the gas until the CO$_2$ snowline, which is unresolved in most data.  
However, the situtation may be more complicated than this as a study of accreting T Tauri stars demonstrates that the general trend is that carbon is depleted in the inner disk, which we define as the gas accretion disk inside the dust sublimation radius, typically at 0.01 - 0.1 AU; \citet{2019A&A...632A..32M} finds depletion factors ranging between 2 and 25 relative to the diffuse ISM value of 1.35 x 10$^{-4}$.
The volatile and refractory depletion trend across different atoms in the well studied inner disk of TW Hya \citep{2019A&A...632L..10B,2020A&A...642L..15M} is the inverse of the super solar abundances observed in solar system meteorites \citep{2010ASSP...16..379L,2015PNAS..112.8965B,2020A&A...642L..15M}, which demonstrates that the observed carbon depletion likely results from the trapping of ice-coated rocky pebbles in the outer disk, either in a pressure maximum or in bodies that grow quickly to kilometer-sized planetesimals stopping the radial drift.
Since some of the disks in \citet{2019A&A...632A..32M} are compact in the dust and do not show any signs of strong pressure bumps, it is proposed that some of these sources have produced planetesimals via the streaming instability, creating an initial dust trap before any planets have carved a gap.
However, the carbon abundance in the outer regions of the disks in the \citet{2019A&A...632A..32M} survey has never been unambiguously determined. Therefore it is still unclear whether the carbon abundance in the outer disk is even lower than the inner disk, as expected based on the models including radial drift, or that the dust is efficiently trapped in dust traps, pebbles and planetesimals.

Observational estimates of the elemental volatile carbon abundance in protoplanetary disks are difficult to constrain.
The main gas-phase carbon carriers in the atmospheric layers of protoplanetary disks are C$^{\rm +}$, C$^{\rm 0}$, CO, and CO$_2$, ordered by increasing shielding from ultraviolet photons \citep{2012A&A...541A..91B,2014FaDi..168...61B,2018A&A...618A.182B,2020ApJ...899..134K}.
CO is the dominant reservoir in molecular gas in the bulk of the disk, but is not the most accurate probe of the C/H abundance as most isotopologues are optically thick \citep[see e.g.][]{2020ApJ...891L..17Z,2020MNRAS.493L.108B} and can be affected by direct photodissociation and freeze-out \citep{2016A&A...588A.108K}. 
In the outer and upper regions of the disk where CO and small dust are depleted and the effects of photodissociation are magnified, neutral and ionized carbon become the main carbon reservoirs in the gas. 
Atomic carbon is not often observed in protoplanetary disks and may have non-disk emission components, but is in combination with spatially resolved CO emission a better tracer of the elemental volatile carbon abundance than CO alone \citep{2016A&A...588A.108K}.

In this work, we present [C~I] observations using the Atacama Compact Array (ACA) of the Atacama Large Millimeter/submillimeter Array (ALMA) in a sample of seven disks, four of which have a known carbon abundance in the inner disk as determined from accreting material emitting in the near-infrared \citep{2019A&A...632A..32M}. 
Three of these disks were identified as having a dust trap on the basis of their inner disk carbon abundances.
The sample, observations, and data analysis are described in Sect. \ref{sec:methods}, the analysis of the observed spectra in Sect. \ref{sec:obeservational_results}.
Using the atomic carbon emission in combination with archival CO data, we determine a radial profile of the elemental carbon abundance in the three sources with known inner disk carbon abundance. 
The modeling methodology is described in Sect. \ref{sec:model_analysis}, the results of the modeling in Sect. \ref{sec:modeling_results} and the implications for the disk chemistry and formation of planetesimals in Sect. \ref{sec:discussion}.

\begin{figure*}[!b]
    \centering
    \includegraphics[width=\linewidth]{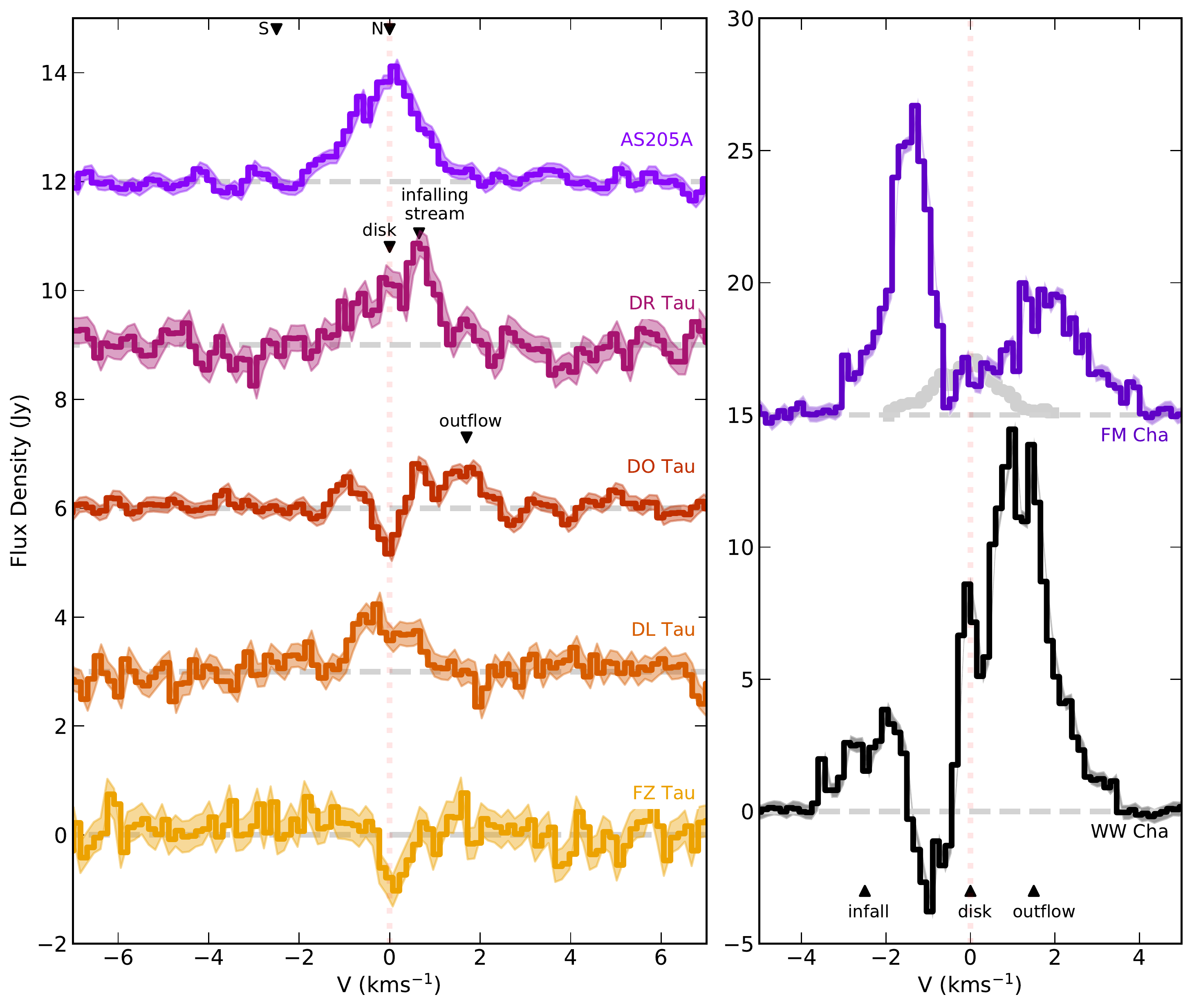}
    \caption{Spatially integrated emission spectra of the [C~I] $J$ = 1-0 line at 492.165 GHz for the seven sources in our sample. All the spectra are shifted to their respective local standard of rest velocity (Table \ref{tab:stellar_props}). The shaded region marks the 1$\sigma$ uncertainty for each source. The spectrum of AS 205 A is shown in grey with FM Cha as a reference of the typical linewidth for a modestly inclined disk. N and S depict the $V_{\rm lsr}$ of AS 205 N and AS 205 S respectively.}
    \label{fig:spectra}
\end{figure*}
\begin{figure*}[ht]
    \centering
    \includegraphics[width = 1\linewidth]{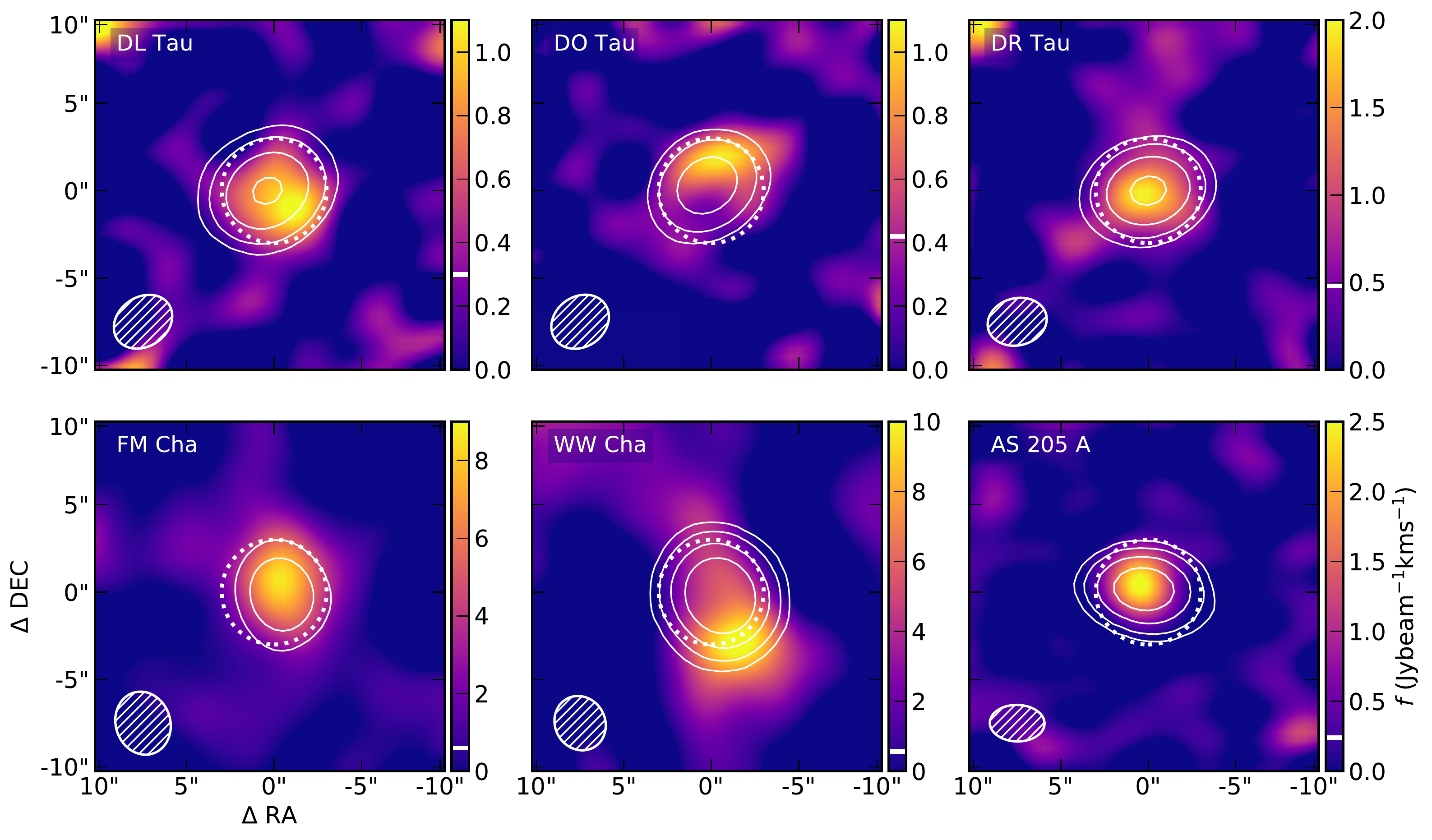}
    \caption{Moment zero maps of the [C~I] $J$ = 1-0 emission. Solid white contours show the 0.61 mm continuum at 10, 30, 100, 300 $\sigma$ respectively. The dashed white contour indicates the region used to extract a spectrum. Beam sizes are shown in the lower left corner of each panel. The 3$\sigma$ detection limit is illustrated as a white tickmark in the colorbar.}
    \label{fig:mom0maps}
\end{figure*}

\section{Observations and data analysis}
\label{sec:methods}
The full sample contains seven well-studied disks, four in Taurus, two in Chamaeleon I, and one in Ophiuchus. 
The sample was selected to have strong near-infrared permitted C$^0$ lines tracing the inner disk in either the McClure (2019) data or archival VLT X-shooter spectra.
Subsequent criteria were that the sources already have CO isotopologue 12m ALMA observations existent from other programs that could provide additional constraints on our modeling.
%This sample helps to discriminate low elemental volatile carbon abundances in typical protoplanetary disks from other CO depleting mechanisms. 
%Until now, TW Hya is the only disk in which carbon depletion is unambiguously determined in the outer disk using multiple carbon carrying species.
Table \ref{tab:stellar_props} lists the properties of the stars in our ALMA sample.
Most of the stellar properties are taken from \citet{2019A&A...632A..32M}.
The rest velocities of the sources are taken from high resolution CO data. 

\subsection{Observations}
\label{ssec:observations}
Our ALMA Atacama Compact Array (ACA) Band 8 observations were conducted in August and September of 2018 as program 2017.1.00857.S (PI: M.K. McClure).
The continuum emission was recorded in two spectral windows (SPWs), centered on
479.59 GHz and 477.7 GHz respectively, each with a bandwidth of 2.0 GHz.
One of the line SPWs was centered on the [C~I] $J$ = 1-0 line at 492.165 GHz, the other line SPW was centered on the CS $J$ = 10-9 line at 489.755 GHz in an attempt to constrain the sulfur chemistry in the outer disk. 
The latter emission line is not detected in any of the sources, and hence it is not used in this analysis.
The bandwidth of the line SPWs were 250 MHz each, resulting in a spectral resolution of 122 kHz or 0.086 km s$^{-1}$.
All disks are unresolved in the beam of the ACA, which is typically 3.5$\dprime$ x 2.7$\dprime$.

The selected sample was split into three observing groups, based on their position on the sky and the integration time that was needed to get to the required sensitivity.
The sources in Taurus were observed in seven different executions for $\sim$1-2 hours integration time per source, the sources in Chamaeleon in five different executions for $\sim$2.5 hours integration time per source and AS 205 A in two different executions for 37 minutes in total. 
Full details of the observations and used calibrators are summarized in Table \ref{tab:observing_log}.

\subsection{Data analysis}
\label{ssec:data_analysis}
The data were reduced with the standard ALMA pipeline in CASA \citep{2007ASPC..376..127M}. 
The sources in Taurus and AS 205 A were calibrated using \texttt{CASA} v5.1.1, the sources in Chamaeleon were calibrated with \texttt{CASA} v5.4.0.
Self-calibration was performed on one of the continuum spectral windows of the combined executions, except for the sources FZ Tau and FM Cha that have a very weak continuum emission (see Table \ref{tab:line_fluxes}). 
The self-calibration tables were then applied to the line SPWs.
For the sources that have high enough $S/N$ to apply self-calibration, this procedure provided an improvement in continuum image quality of 50-100\% and improved the line peak $S/N$ by $\sim10$\%.
In order to gain $S/N$ in the [C~I] $J$ = 1-0 line, especially for the sources with weak line emission, we rebinned the Measurement Set using the $\texttt{cvel2}$ command in \texttt{CASA}, reducing the velocity resolution by a factor of $\sim$2 to 0.15 km s$^{-1}$.

The continuum is subtracted in the uv-plane after de-selection of the channels with line emission based on the dirty image.
All sources, except FZ Tau where no [C~I] $J$ = 1-0 emission was detected, were imaged using the $\texttt{tclean}$ task in \texttt{CASA} with Briggs weighting and a robust factor of 0.5.
The choice of this robust factor is a trade-off between high sensitivity for the weak lines and a small beam to maximize the ability to discriminate between disk and cloud emission.
The emission and absorption features were masked during the imaging procedure using hand-drawn regions in each channel.
The line spectra were extracted using a circular mask with a radius of 2.5$\dprime$, centered on the continuum peak, and are plotted in Fig. \ref{fig:spectra}.
Continuum fluxes, continuum RMS noise and the final beam size of each source are listed in Table \ref{tab:observing_log}.
The disk component of the line flux of the [C~I] $J$ = 1-0 emission was selected based on literature $V_{\rm lsr}$ values taken from CO emission (Table \ref{tab:stellar_props}) and the modeling.

\subsection{Complementary data}
\label{ssec:complementary_data}
In addition to the [C~I] mini-survey, we analyzed archival ALMA 12m array data of $^{13}$CO and C$^{18}$O emission for the three sources that we modeled and studied in more detail: DR Tau, DO Tau and DL Tau.
The details of the different molecular transitions can be found in Table \ref{tab:lines}.
The Band 6 CO $J$ = 2-1 isotopologue data were observed as part of program 2016.1.01164.S (PI: G. J. Herczeg). 
Details of the data and initial calibration can be found in \citet{2019ApJ...882...49L}.
The $^{13}$CO and C$^{18}$O $J$=2-1 is imaged using natural weighting, with a typical beam-size in the channel maps of 0.14$\dprime$ x 0.11$\dprime$ at a velocity resolution of 0.2 km s$^{-1}$.
Spectra were extracted using a circular aperture with a radius of 1.5$\dprime$ for DL Tau and DR Tau and 1$\dprime$ for DO Tau.

For DR Tau we included additional Band 7 CO $J$ = 3-2 isotopologue data, as part of program 2016.1.00158.S (PI: E.F. van Dishoeck).
This dataset was calibrated using \texttt{CASA} v4.7.3. After the standard pipeline calibration, we performed self-calibration on the continuum and applied the gain solutions to the line channels.
The beam-size in the final images is 0.24$\dprime$ x 0.20$\dprime$ at a velocity resolution of 0.055 km s$^{-1}$. 
Self-calibration on the continuum reduced the noise by a factor of 2 to 0.13 mJy/beam. 
The total source integrated continuum flux is 333.7 $\pm$ 0.2 mJy.
The spectrum is extracted using a circular aperture with a radius of 2$\dprime$.

\begin{table}[ht]
    \caption{Molecular transitions presented in this paper.}
    \label{tab:lines} 
    \begin{tabular}{L{.2\linewidth}C{.2\linewidth}C{.2\linewidth}C{.2\linewidth}}
    \hline\hline
    Transition \Tstrut& $\nu$ \Tstrut & $E_\mathrm{u}$  \Tstrut& $A_{\mathrm{ul}}$ \Tstrut\\
    & (GHz)\Bstrut&(K)\Bstrut&( s$^{-1}$)\Bstrut\\
    \hline
    [C~I] $J$ =1-0 \Tstrut&\Tstrut 492.1607 &\Tstrut 23.6 &\Tstrut 7.88 x 10$^{-8}$\\
    $^{13}$CO $J$=2-1 & 220.3987 & 15.9 & 6.08 x 10$^{-7}$\\
    $^{13}$CO $J$=3-2 & 330.5880 & 31.7 & 2.19 x 10$^{-6}$\\
    C$^{18}$O $J$=2-1 & 219.5604 & 15.8 & 6.01 x 10$^{-7}$\\
    C$^{18}$O $J$=3-2 & 329.3306 & 31.6 & 2.17 x 10$^{-6}$\\    
    \hline\noalign {\smallskip}
    \end{tabular}
    \noindent \textbf{Notes.} The frequency, energy levels and Einstein $A$ values are adopted from the CDMS database \citep{2005JMoSt.742..215M}.
\end{table}    

\section{Observational results}
\label{sec:obeservational_results}
\begin{table*}[!t]
    \caption{Observed [C~I] $J$=1-0 line and continuum fluxes}
    \label{tab:line_fluxes}
    \begin{tabular}{L{.1\linewidth}L{.11\linewidth}L{.11\linewidth}L{.11\linewidth}L{.09\linewidth}L{.11\linewidth}L{.09\linewidth}L{.09\linewidth}}
    \hline\hline \noalign {\smallskip}
    Source& Total Line Flux &Integration limits & Flux Disk Component &Disk Peak $S/N$&Line Channel RMS&Continuum Flux&Continuum $S/N$\\
    &(Jy km s$^{-1}$)&(km s$^{-1}$)&(Jy km s$^{-1}$) & & (Jy)&(mJy)&\\
    \hline\noalign {\smallskip}
    DL Tau & 1.36 $\pm$ 0.10 &-1.0 - 1.0& 1.36 $\pm$ 0.10 & 5.59&0.18  &889    $\pm$ 2.3   &387\\
    DO Tau &1.33 $\pm$ 0.14 &-0.9 - 3.1& 1.18 $\pm$ 0.10 & 7.79&0.19   &529    $\pm$ 2.5   &212\\
    DR Tau &2.68 $\pm$ 0.16 &-1.8 - 1.9& 0.92 $\pm$ 0.16 & 4.57&0.23   &1008   $\pm$ 2.6   &388\\
    FZ Tau &< 0.87 & - &-&-&0.21                                &51     $\pm$ 2.8   &18\\
    AS 205 A &3.42 $\pm$ 0.08&-2.1 - 2&3.42 $\pm$ 0.12&17.06&0.11    &1793   $\pm$ 2.4   &747\\
    FM Cha &23.65 $\pm$ 0.20&-3.2 - 3.7& - & - &0.20                   &204    $\pm$ 2.8   &73\\
    WW Cha &30.27 $\pm$ 0.19&-3.6 - 3.4& - & - &0.18                   &2662   $\pm$ 2.4   &1109\\
    \hline
    \end{tabular}
    \\
    \Tstrut
    \noindent \textbf{Notes.} All uncertainties are given as 1$\sigma$, upper limits are given as 3$\sigma$. Peak $S/N$ is given for the disk component, line RMS is given for the final images over a frequency width of 0.15 kms$^{-1}$.
    \end{table*}
Six out of the seven sources are detected in [C~I] with high confidence. 
All spectra are presented in Fig. \ref{fig:spectra}. Emission is seen to be within -2 and 2 km s$^{-1}$ for the sources in Taurus and AS 205 A and within -4 and 4 km s$^{-1}$ in the sources in Chamaeleon I. Therefore, we use all channels with a flux density above zero within these limits to construct the moment 0 maps as shown in Fig. \ref{fig:mom0maps}. The integration limits for each source are given in Table \ref{tab:line_fluxes}.
In most sources where [C~I] emission is detected, the emission is either extended or suffers from foreground absorption.
Integrated line fluxes, as well as the disk contribution, are listed in Table \ref{tab:line_fluxes}.
In the forthcoming section, we describe the emission of each source separately.\\

\textbf{DR Tau} is a relatively young source \citep[0.9 Myr,][see Table \ref{tab:stellar_props}]{2019A&A...632A..32M} with a compact, almost face-on disk \citep[dust $R_{\mathrm{out}} = 54$ AU, $i$~=~5.4~$^\mathrm{o}$,][]{2019ApJ...882...49L}, that has no signs of structure in the millimeter continuum emission image \citep{2019ApJ...882...49L}.
In the [C I] $J$ = 1-0 spectrum we can identify three different emission components.
One emission peak, centered at the cloud velocity, is emission associated with the disk.
This emission is single-peaked as expected based on the low inclination.
The disk component is spaced over 5 channels, which means that the emission is spectrally resolved.
The disk has an integrated line flux of 0.92 $\pm$ 0.16 Jy km s$^{-1}$.

On the redshifted side (centered at 0.7 km s$^{-1}$) of the [C~I] $J$ = 1-0 spectrum, we detect an additional emission component from a stream of material that is likely infalling. We discuss this extra component further in Sect. \ref{sec:modeling_results}.
%The $^{13}$CO emission has a small excess at the same relative velocity, much weaker in comparison to the signal from the disk (see Fig. \ref{fig:dr_lines}).
%This non-Keplerian excess seems to trace a gas filament infalling on the disk, exposed to a high UV flux which yields the high [C~I]/$^{13}$CO emission.
Some additional emission is observed on the blueshifted side of the spectrum.
The [C~I] channel maps at this velocity show small extended emission and absorption features (see Fig. \ref{fig:drtaucube}), but the spatial resolution is insufficient to see clear structure in the emission.\\

\textbf{DO Tau} is a 2 Myr old source (see Table \ref{tab:stellar_props}) harbouring a compact dust disk with an outer radius of 37 AU and an inclination of 27.6$^{\rm o}$, without any substructure in the millimeter continuum emission image \citep{2019ApJ...882...49L}. 
This source has a clear detection in [C~I] $J$ = 1-0, with two peaks separated by an absorption feature at cloud velocity.
This absorption feature can be caused by the presence of extended emission in the data beyond the maximum resolvable scale of the interferometer, foreground cloud absorption or a combination of both.
The redshifted non-Keplerian emission centered at 1.5 km s$^{-1}$ is connected with extended emission in the channel maps that appears to be co-located with the blueshifted part of the pole-on outflow observed in $^{12}$CO \citep{2020AJ....159..171F}, as illustrated in Appendix \ref{app:diskwind_DO}.
The moment zero map of DO Tau shows very little emission, due to mostly the foreground cloud.
For DO Tau we determine two different fluxes, dealing differently with the absorption feature. 
The integrated emission of the disk contribution is 0.59 $\pm$ 0.10 Jy km s$^{-1}$, masking the absorption feature.
Assuming a constant flux of 0.6 Jy at velocities where we observe cloud absorption results in an integrated flux of 1.18 $\pm$ 0.10 Jy km s$^{-1}$. 
The latter approach is more useful in the modeling and is justified based on our modeling and the clean line profiles of the other two sources that have an inclination similar or higher than DO Tau (AS 205 A and DL Tau).\\

\textbf{DL Tau} is an old source \citep[7.8 Myr,][ we discuss its age more specifically in Sect. \ref{ssec:comparison_modeling}]{2019A&A...632A..32M} and has a large continuum disk with a 95\% effective radius of 147 AU, an inclination of 27.6 $^{\rm o}$ and three dust rings \citep{2018ApJ...869...17L}.
On-source [C~I] emission is detected without any signs of extended emission in the clean channel maps (see Fig. \ref{fig:dltaucube}). 
The spectrum reveals a slightly asymmetric double-peaked Keplerian profile, with the blueshifted peak slightly brighter than the redshifted emission.
The integrated line flux of DL Tau is 1.36 $\pm 0.10$ Jy km s$^{-1}$.
Even though the source has a large dust disk, no CO isotopologue emission is detected, which suggests that this source is highly depleted in CO.\\

\textbf{FZ Tau} is a 0.9 Myr old source (see Table \ref{tab:stellar_props}) with a small \citep[$\sim$10 AU][]{2014A&A...564A..95P} dust disk and an inclination of <70~$^{\rm o}$. 
FZ Tau has no detected [C~I] $J$ = 1-0 emission, but does show a similar absorption feature as DO Tau.
The angular distance between FZ Tau and DO Tau is small (2.3~$^\mathrm{o}$), which means that the observed absorption is probably extended cloud absorption.
The most likely reason for the non-detection is the small disk size \citep[$\sim$10 AU based on the continuum][]{2014A&A...564A..95P}.
FZ Tau has strong near-infrared permitted C$^0$ emission \citep{2019A&A...632A..32M} that requires a large enough gas-rich hot region interior to the dust sublimation radius and sufficient UV to photodissociate CO. 
Unfortunately, the beam dilution is too large to detect the [C~I] emission tracing the outer disk using ACA.\\

\textbf{WW Cha} is a 6.4 Myr old source (see Table 1) with a 135 AU structured dust disk at an inclination of 37.2 $^{\rm o}$ \citep{2021ApJ...909..212K}. 
WW Cha has a [C~I] spectrum with a classic P-Cygni profile, indicative of an outflow, consisting of a broad intense emission feature on the redshifted side of the spectrum together with a narrower and a weaker absorption feature in the blueshifted emission.
Besides the outflow we detect a disk component near the cloud velocity and a ridge of infalling material on the blueshifted side of the spectrum (see also the channel maps in Fig. \ref{fig:wwchacube}).
The source is known to have a jet \citep{2012AJ....144...83R} and lately confirmed, using VLTI, to be a close binary with a separation of 1.01 AU \citep{2015A&A...574A..41A,2021arXiv210200122G}.
The P-Cygni profile could be the result of an outflow caused by the binary interactions with the disk.\\

\textbf{FM Cha} is 5.3 Myr old source (see Table 1) that harbours a 82 AU dust disk at an inclination of 51 $^{\rm o}$ \citep{2020ApJ...895..126H}. 
FM Cha has a firm detection in [C~I] $J$ = 1-0 with two asymmetric peaks in the spectrum on either side of the systemic velocity. 
The [C~I] line in FM Cha is much broader than in AS 205 A, which we add in grey in Fig. \ref{fig:spectra} for reference, even though both disks have a similar inclination.
Most of the emission is extended, contaminated cloud emission at high velocities (>1 km s$^{-1}$) relative to the local standard of rest.
Although we observe some hints of Keplerian emission close to the source velocity, we cannot constrain the disk contribution in the emission due to the low spatial resolution of the [C~I] observations.\\

\textbf{AS 205 A} is a binary system with a large, 53 AU, dust disk to the north (AS 205 N) that is with an inclination of 20.1~$^\mathrm{o}$ relatively face-on, and a small, 24 AU, dust disk to the south (AS 205 S) that is inclined at 66.3~$^{\rm o}$ \citep{2018ApJ...869L..44K}.
The system shows a lot of substructures in CO, due to gravitational interaction between the two systems.
The distance between the two disks is 1.3$\dprime$, which means that the binary system is unresolved in the beam of our observations.
The neutral carbon emission is singly peaked, centered on the systemic velocity of the largest disk AS 205 N.
The systemic velocity of the secondary star is offset by -2.5 km s$^{-1}$ in CO \citep{2018ApJ...869L..44K}, as indicated in Fig. \ref{fig:spectra}.
There is no significant emission detected at this relative velocity, which means that AS 205 S is not detected and that it is reasonable to assume that all the observed emission comes exclusively from AS 205 N.\\

Concluding, four disks out of the seven targets have been detected in [C~I] $J$=1-0 in this study.
In the following we describe the modeling of the three Taurus disks in the sample for which the [C~I] $J$=1-0 emission can be easily decomposed. 
These three sources already have well-determined inner disk C abundance in the literature, which makes them more suitable to study the radial profile of the carbon abundance.

\section{Description of the modeling}
\label{sec:model_analysis}
The aim of our modeling was to develop a physical disk model tailored to each source that describes both the dust and the gas structure in each disk.
Using this model, we could then determine the volatile carbon abundance in the outer disk by comparing the flux in the [C~I] $J$ = 1-0 observations with a grid of models varying carbon and oxygen abundances.
In order to determine the physical structure of the disks, we combined information from the spectral energy distribution (SED) and the outer radii of the disks in ALMA millimeter continuum and CO observations.
Although the model may not be unique, it constrains a structure that agrees well with the observations.
See \citet{2016A&A...588A.108K} for a detailed overview of the impact of changes in variables of the physical disk model on the line flux in [C~I] $J$ = 1-0.

The azimuthally symmetric physical-chemical model DALI \citep[Dust And LInes;][]{2012A&A...541A..91B,2013A&A...559A..46B} is used to model the sources using a smooth parameterized gas and dust disk density structure, neglecting all azimuthal substructures, gaps and rings seen in the dust. 
DL Tau is known to have multiple gaps in the millimeter continuum \citep{2018ApJ...869...17L}, but the influence of the millimeter dust on the gas temperature in the outer disk is small \citep[see e.g.,][]{2018A&A...612A.104F,2018ApJ...867L..14V,2020ApJ...905...68A}, so it is safe to use a smooth disk density profile to determine the gas temperature and chemistry in the outer disk.
Both DO Tau's and DR Tau's disk are smooth in the millimeter dust.
DALI first solves the radiative transfer of the continuum based on a dust density structure (Sect. \ref{ssec:disk_parameters}) and an input external radiation field (Sect. \ref{ssec:stellar_parameters}) to determine the dust temperature and radiation field from UV to millimeter wavelengths at all locations in the disk. 
After that, the heating-cooling balance of the gas is solved simultaneously with steady state chemistry in an iterative sequence until convergence is reached.
Once the solution is converged, we ray-trace the model in continuum and molecular lines using non-LTE radiative transfer (Sect. \ref{ssec:radiative_transfer}). 

\subsection{Disk parameters}
\label{ssec:disk_parameters}
The density structure of our model is fully parameterized and is based on a full disk, without inner cavity.
We used the standard form of the dust surface density profile of a power law with an outer exponential taper:
\begin{equation}
\Sigma_{\mathrm{dust}}=\Sigma_{\mathrm{c}} \left(\frac{r}{R_{\mathrm{c}}}\right)^{-\gamma} \exp \left[-\left(\frac{r}{R_{\mathrm{c}}}\right)^{2-\gamma}\right],
\end{equation}
defined by the surface density $\Sigma_{\mathrm{c}}$ at the characteristic radius $R_\mathrm{c}$, and assuming an initial power-law index $\gamma$ = 1.
The scale height $h$ at distance $r$ is given by
\begin{equation}
    h=h_{\mathrm{c}}\left(\frac{r}{R_{\mathrm{c}}}\right)^{\psi},
\end{equation}
defined by the scale height $h_{\mathrm{c}}$ at the characteristic radius with a flaring index $\psi$.
The dust was split into two populations, following the approach of \citet{2006ApJ...638..314D}, a small dust population which ranges from 0.005 to 1 micron and a large dust population which ranges from 0.005 to 1000 micron. 
The vertical structure of the small dust grains is 

\begin{equation}
    \label{eq:small_grains_distribution}
    \rho_{\mathrm{d}, \mathrm{small}}=\frac{(1-f) \Sigma_{\mathrm{dust}}}{\sqrt{2 \pi} r h} \exp \left[-\frac{1}{2}\left(\frac{\pi / 2-\theta}{h}\right)^{2}\right],
\end{equation}
where $f$ is the mass fraction of the large grain distribution, assumed to be 0.99, and $\theta$ is the opening angle from the midplane as seen from the central object.
The large grains are settled with respect to the full disk height with a factor $\chi \in (0,1]$, following the same distribution as the small grains (Eq. \ref{eq:small_grains_distribution}) but with $f$ replacing (1-$f$) and $\chi \times h$ replacing $h$.
The radial distribution of the two dust grain populations is the same, excluding grain evolution processes.
The gas density distribution is scaled up from the dust density distribution using the ISM gas-to-dust ratio of 100.
The vertical gas structure follows the distribution of the small grains, preserving the global gas-to-dust ratio.
We make an initial estimate of the sublimation radius in the disk using Equation 14 in \citet{2001ApJ...560..957D}. 
By assuming a dust sublimation temperature of 1500 K, negligible self-irradiation ($H_{\rm rim}$/$R_{\rm rim}$ << 1), and solar units for the luminosity, Equation 14 reduces to $r_{\rm subl}$ [AU] = 0.07 $(L[L_{\odot}])^{1/2}$ \citep[see also][]{2002ApJ...579..694M}.
%An initial guess for the sublimation radius in the disk is made using equation 14 in \citet{2001ApJ...560..957D}, assuming a dust sublimation temperature of 1500 K and negligible self-irradiation: $r_\mathrm{subl}$ =  0.07$\sqrt{L/\mathrm{L_\odot}}$ AU.
For DO Tau we increased this result to 0.14 AU to match the mid-infrared emission from the inner disk, similar to the value found for the Br$\gamma$ area for DO Tau as determined in \citet{2014MNRAS.443.1916E}.

\subsection{Stellar parameters}
\label{ssec:stellar_parameters}
The stellar spectra are approximated by a perfect blackbody defined by the effective temperature ($T_\mathrm{eff}$) of each star with an additional component due to accretion, assuming an accretion temperature of 10,000 K of the emitting radiation, using 
\begin{equation}
    L_{\mathrm{acc}}(v)=\pi B_{v}\left(T_{\mathrm{acc}}, v\right) \frac{G M_{*}\dot{M}}{R_{*}} \frac{1}{\sigma T_{\mathrm{acc}}^{4}}.
\end{equation}
The observed far infrared luminosity of the sources \citep{2012ApJ...744..121Y} agree  within an order of magnitude with this first order approximation. 

The properties of all modeled stars are taken from \citet{2019A&A...632A..32M}, which uses the procedures from \citet{2013ApJ...769...73M} to derive self-consistent stellar and accretion parameters.
We included a X-ray luminosity of 10$^{30}$ erg s$^{-1}$ at a X-ray plasma temperature of 7 x $10^{7}$ K, and a cosmic ionization rate of 5 x $10^{-17} \mathrm{s}^{-1}$ incident on the disk surface.

\subsection{Chemical network}
\label{ssec:chemical_network}
The standard chemical network adopted in DALI is based on UMIST06 \citep{2007A&A...466.1197W}.
This chemical network includes neutral and charged PAHs, 109 molecular species and 1463 reactions.
The code includes two-body interactions, freeze-out, photodissociation and -ionization, and thermal- and photo-desorption.
There are no grain surface reactions considered, except for the hydrogenation process.
It is reasonable to assume that this has no impact on the species that are used in this paper, which trace the atmospheric layers of the disk.
For DR Tau and DO Tau we include the full treatment of CO freeze-out and isotope selective processes as described by \citet{2014A&A...572A..96M,2016A&A...594A..85M}.
We adopt fiducial ISM gas-phase elemental carbon and oxygen abundances of C/H = 135 ppm and O/H = 288 ppm, respectively. 
These values are close to the median abundances observed in translucent interstellar clouds \citep{1996ApJ...467..334C,1998ApJ...493..222M,2012ApJ...760...36P} and consistent with the solar values within a factor of a few \citep{2009ARA&A..47..481A}. 
See \citet{2012A&A...541A..91B} for a full justification of these fiducial abundances.
In order to make the carbon abundances easy to interpret, we define a carbon depletion factor ($f_\mathrm{depl}$) as (C/H)$_\mathrm{ISM}$/(C/H).
When changing the carbon abundance we vary the oxygen abundance such that the C/O ratio stays at the ISM level of 0.47, unless specified differently.
Without a similar depletion of oxygen w.r.t. carbon, the neutral atomic carbon abundance would be more strongly dependent on the elemental carbon abundance, as there would be more oxygen available to create CO.

\subsection{Radiative transfer}
\label{ssec:radiative_transfer}
After the chemical network has converged to a steady state equilibrium and gas temperature solution, atomic and molecular transitions can be imaged using the ray-tracer in DALI which includes the radiative transfer.
The excitation of the line is determined explicitly in a non-LTE fashion, using the collision rate coefficients from the LAMDA database \citep{2005A&A...432..369S,2020Atoms...8...15V}.
In this step, the sources are projected on the sky using the observed distance, inclination, and position angle given in Table \ref{tab:stellar_props} and Table \ref{tab:model_params}, respectively.
For comparison with our observations, we convolved the model with a Gaussian beam with the same shape as the beam of the ALMA observations.

\begin{figure*}
    \centering
    \includegraphics[width=\textwidth]{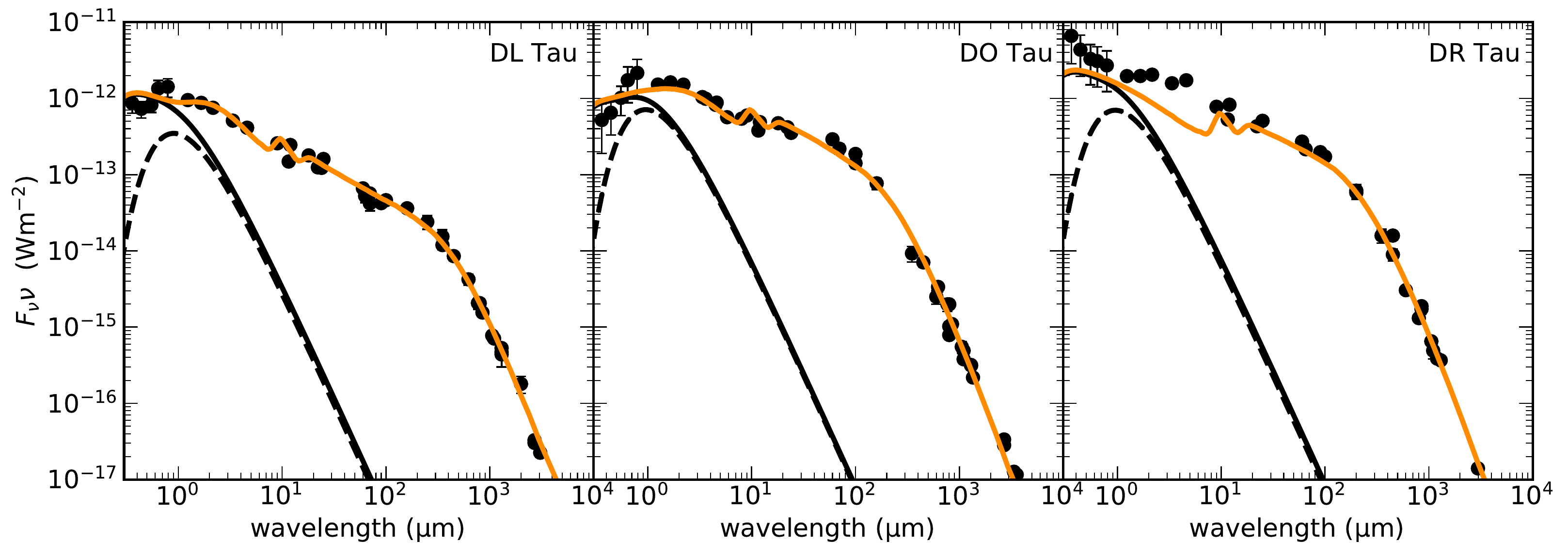}
    \caption{SEDs of the three modeled sources with our model fit in orange. The black dots with errorbars represent observations, the dashed black curve denotes the stellar component, the solid black curve is a combination of the stellar component and the UV excess due to accreting material as used in the modeling.}
    \label{fig:seds}
\end{figure*}

\subsection{Fitting procedure}
In the fitting procedure, we followed a similar approach as \citet{2014A&A...562A..26B}, creating a physical model that fits the SED through consecutive parameter value refinements, until they converge to a model that is best representative of the source.
Using a $\chi^2$ or Markov chain Monte Carlo (MCMC) method was not possible because of the long computation time of the continuum radiative transfer.
We used the same characteristic radius ($R_{\rm c}$), inclination ($i$) and position angle (PA) for the gas as observed in high angular resolution dust continuum observations. 
The gas disk can be significantly larger than expected from millimeter continuum observations if radial drift and grain growth are important in the disk \citep{2017A&A...605A..16F,2019A&A...629A..79T}, but for the sake of simplicity we do not include dust evolution in our modeling.
The high spatial resolution sub-millimeter continuum constrains the mass, the characteristic radius ($R_{\rm c}$), the inclination ($i$) and the position angle (PA) of the dust in the disks. 
Additionally, we look at the spectral line profiles of the CO isotopologue and [C~I] emission to further constrain the density and temperature structure of the disks.
These properties together constrain $\Sigma_{\rm dust}$, $\gamma$, $\chi$, $f$, $h_{\rm c}$ and $\psi$.

\subsection{Main sources of uncertainty}
We stress that the gas phase carbon abundances found in the modeling analysis are subject to a few uncertainties. 
First, we assume a smooth disk with a system averaged elemental volatile carbon depletion.
It is thought that most CO processing preferentially occurs at the deepest, coldest layers \citep{2018A&A...618A.182B}, while the tracers that we analyze, such as CO and C$^{0}$ are known to trace the atmospheric layers of the disk \citep[$z/r$ > 0.2][]{2012A&A...541A..91B}.
Therefore, both radial and vertical dependence in (C/H)$_{\rm gas}$ can cause discrepancies in the abundance found in the system averaged CO isotopologue and [C~I] emission.
Based on the carbon abundance measured in the upper layers of the disk, we infer that the bulk of the disk can be depleted in carbon, which is a result from efficient mixing by turbulence.
Additionally, abundances of carbon carrying molecules depend on the elemental oxygen abundance in the disk, since an excess in carbon atoms can shift the balance between CO$_2$, CO, C, and hydrocarbon abundances. 
Elevated C/O ratios are not uncommon in disks and can arise by a stronger depletion of oxygen, because of freeze-out and processing of CO$_2$ and H$_2$O rather than CO \citep[which has been inferred directly from water observations, see][]{2010A&A...521L..33B,2011Sci...334..338H,2017ApJ...842...98D} or direct photo-ablation of carbon rich grains \citep{2021ApJ...910....3B}.
We assumed ISM C/O ratio where possible and discuss a potential elevated C/O ratio in Sect. \ref{ssec:results_DO_Tau}.

Second, there is an uncertainty in the (C/H)$_{\rm gas}$ found in these three systems, resulting from the uncertainty in the modeling parameters.
The biggest sources of uncertainty are the disk mass, $R_{\rm c}$ and the gas-to-dust ratio \citep{2016A&A...588A.108K}.
There are no HD observations of the sources in our sample, which means that there is high uncertainty on the gas disk mass and the gas-to-dust ratio.
Based on the few sources that are observed in HD, we do not have reasons to assume that the gas-to-dust ratio in the bulk of the disk is much different from the ISM value of 100 \citep{2016ApJ...831..167M,2013Natur.493..644B}. 
\citet{2016A&A...588A.108K} show that an underabundance of gas-phase carbon by an order of magnitude, such as found here, cannot be easily masked by a choice of parameters. 
Larger depletions will be even more reliably identified. 

\begin{table}[!t]
    \caption{Model parameters}
    \label{tab:model_params} 
    \begin{tabular}{L{.22\linewidth}C{.2\linewidth}C{.2\linewidth}C{.2\linewidth}}
    \hline\hline
    &DL Tau\Tstrut\Bstrut & DO Tau \Tstrut\Bstrut& DR Tau\Tstrut\Bstrut\\
    \hline
    $R_{\mathrm{subl}}$ (AU)\Tstrut&0.06 \Tstrut&0.13 \Tstrut&0.074 \Tstrut\\
    $\gamma$&1.0&1.0&0.5\\
    $\chi$,$f$&0.7, 0.99&0.6, 0.99&0.5, 0.99\\
    $R_\mathrm{c}$ (AU)&112.2 &34  &52 \\
    $\Sigma_\mathrm{c}$ (g cm$^{-2}$)&10 &17 &20 \\
    $h_\mathrm{c}$,$\psi$ &0.14, 0.07&0.2, 0.14&0.135, 0.1\\
    $\Delta_\mathrm{gas/dust}$ &100&100&100\\
    {\tiny Gas mass} (M$_\odot$)&8.9 x 10$^{-2}$&1.4 x 10$^{-2}$ &2.5 x 10$^{-2}$\\
    $L_\mathrm{X}$ (erg s$^{-1}$) &10$^{30}$ &10$^{30}$ &10$^{30}$\\
    $T_\mathrm{X}$ (K)& 7.0 x 10$^{7}$&7.0 x 10$^{7}$&7.0 x 10$^{7}$\\
    $\zeta_{cr}$ ( s$^{-1}$)\Bstrut&5.0 x 10$^{-17}$ \Bstrut&5.0 x 10$^{-17}$ \Bstrut&5.0 x 10$^{-17}$ \Bstrut\\
    \hline
    $i$ (deg) \Tstrut&45\Tstrut&27.6\Tstrut&5.4\Tstrut\\
    $\mathrm{PA}$ (deg) \Bstrut &52.1 \Bstrut&170 \Bstrut&3.4 \Bstrut\\
    \hline
    \end{tabular}
    \\
    \Tstrut
    \textbf{Notes.} Inclination and Position angle of the sources are adopted from high resolution millimeter continuum observations \citep{2019ApJ...882...49L}.
\end{table}    

\section{Modeling results}
\label{sec:modeling_results}
Our best-fit results to the SED for the dust density structure are presented in Figure \ref{fig:seds}.
All data points in the SEDs are dereddened using the CCM89 extinction curve \citep{1989ApJ...345..245C} for sources with $A_V < 3$ (DR Tau and DL Tau). 
For DO Tau we used the extinction curve for $0.3 \leq A_k < 1$ from \citet{2009ApJ...693L..81M} to correct the emission at wavelengths shorter than 25 micron. 
This source has $A_V$ = 3.6 magnitudes \citep[][see Table \ref{tab:stellar_props}]{2019A&A...632A..32M}, more than the range over which the CCM89 extinction curve is accurate.
Reasonable fits are found for the SED, including spatial information from the sub-millimeter continuum.
The inferred disk gas masses (Table \ref{tab:model_params}) are within a factor of a few of the estimated disk mass based on multi-wavelength continuum observations and assuming a gas-to-dust ratio of 100. 
See for example \citet{2015ApJ...808..102K} for DL Tau ($4.1 \pm 0.5 \times 10^{-2}$ M$_\odot$) and DO Tau ($1.4 \pm 0.1 \times 10^{-2}$ M$_\odot$), and \citet{2016A&A...588A..53T} for DR Tau ($1.4 \pm 0.4 \times 10^{-2}$ M$_\odot$).

The mid-infrared part of the SED of DR Tau is underestimated by the best representative model.
Varying $h_\mathrm{c}$ between 0.05 - 0.3, $\psi$ between 0.05 - 0.2 and $R_\mathrm{subl}$ between 0.05-0.3 AU, we find no plausible configuration of the parameters that agrees with the inner disk contribution of the SED <10 micron.
The strong mid-infrared excess in the SED could be caused by an inner disk gas component that is optically thick to its own radiation, producing continuum emission.
Evidence for a gas contribution to the continuum interior to the dust sublimation radius in similar systems can be found in many studies \citep[see e.g.,][]{2008A&A...489.1157K,2014A&A...568A..91M,2017A&A...599A..85L}.
Since we focus on outer disk tracers, we do not expect the accreting gas column to have a strong influence on the results.
Note that our lack of a fit of the inner disk's contribution to the continuum increases the model parameter space for fitting the SED at longer wavelengths.
\begin{figure}[!t]
    \centering
    \includegraphics[width = 1\linewidth]{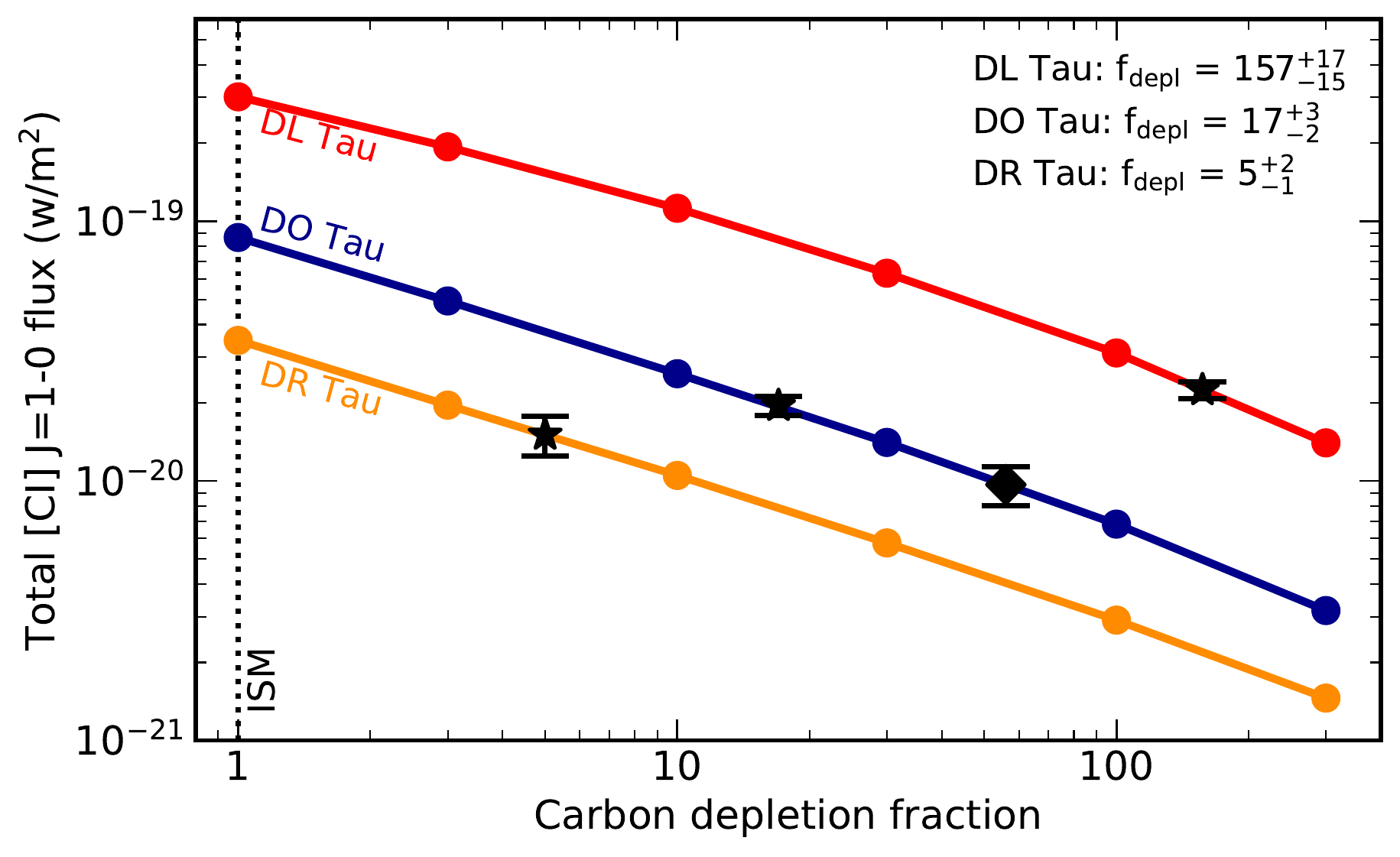}
    \caption{The modeling results of the total flux as function of the carbon depletion factor defined as $f_{\rm depl}$ = (C/H)/(C/H)$_{\rm ISM}$. Observations are plotted at the interpolated carbon depletion fraction. For DO Tau we plot two values for the total flux, the star is the corrected flux for absorption with a constant value, the diamond is only masked for absorption as discussed in the main text. The derived carbon depletion fraction by interpolation with the model is presented in the top-right corner with errors given by measurement uncertainty of the [C~I] observations.}
    \label{fig:chgrid}
\end{figure}

Figure \ref{fig:chgrid} presents the results of the total [C~I] $J$ = 1-0 line flux as function of the elemental gas phase carbon abundance for the three models.
Observations are  presented as black stars with their corresponding depletion factor and uncertainty in the figure legend.
Model carbon abundances are presented as dots at depletion factors of [1,3,10,30,100,300].
The derived depletion factor of the sources is the interpolation of the connecting curves with the error set by the measurement error on the [C~I] $J$=1-0 flux.
A decrease by a factor of ten in the elemental C abundance results typically in a factor of three in the total line flux.
The CO emission and spectra will be discussed source individually in the forthcoming sections.

\subsection{DR Tau: moderate depletion of carbon and oxygen}
\begin{figure}[!t]
    \centering
    \includegraphics[width = 1\linewidth]{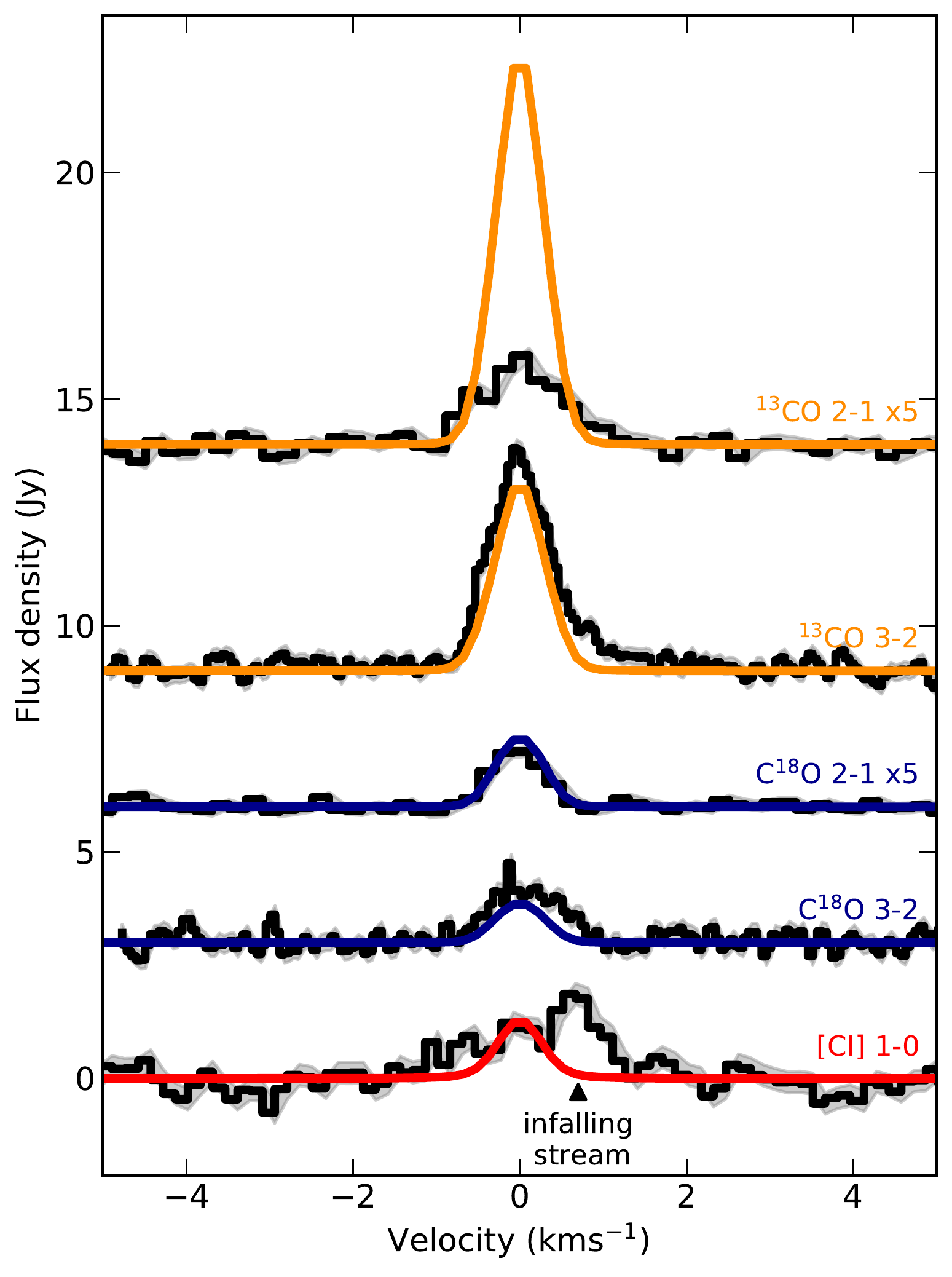}
    \caption{Emission line profiles of $^{13}$CO $J$ = 2-1, $^{13}$CO $J$ = 3-2, C$^{18}$O $J$ = 2-1, C$^{18}$O $J$ = 3-2, and [C~I] $J$ = 1-0 for DR Tau, with the best fitted model overlaid in color. 1$\sigma$ errorbars are indicated in grey for each spectrum. The data and models of the weak CO isotopologue $J$=2-1 transitions have been multiplied by a factor of 5 for visualization purposes.}
    \label{fig:dr_lines}
\end{figure}

\begin{figure}[!t]
    \centering
    \includegraphics[width = 1\linewidth]{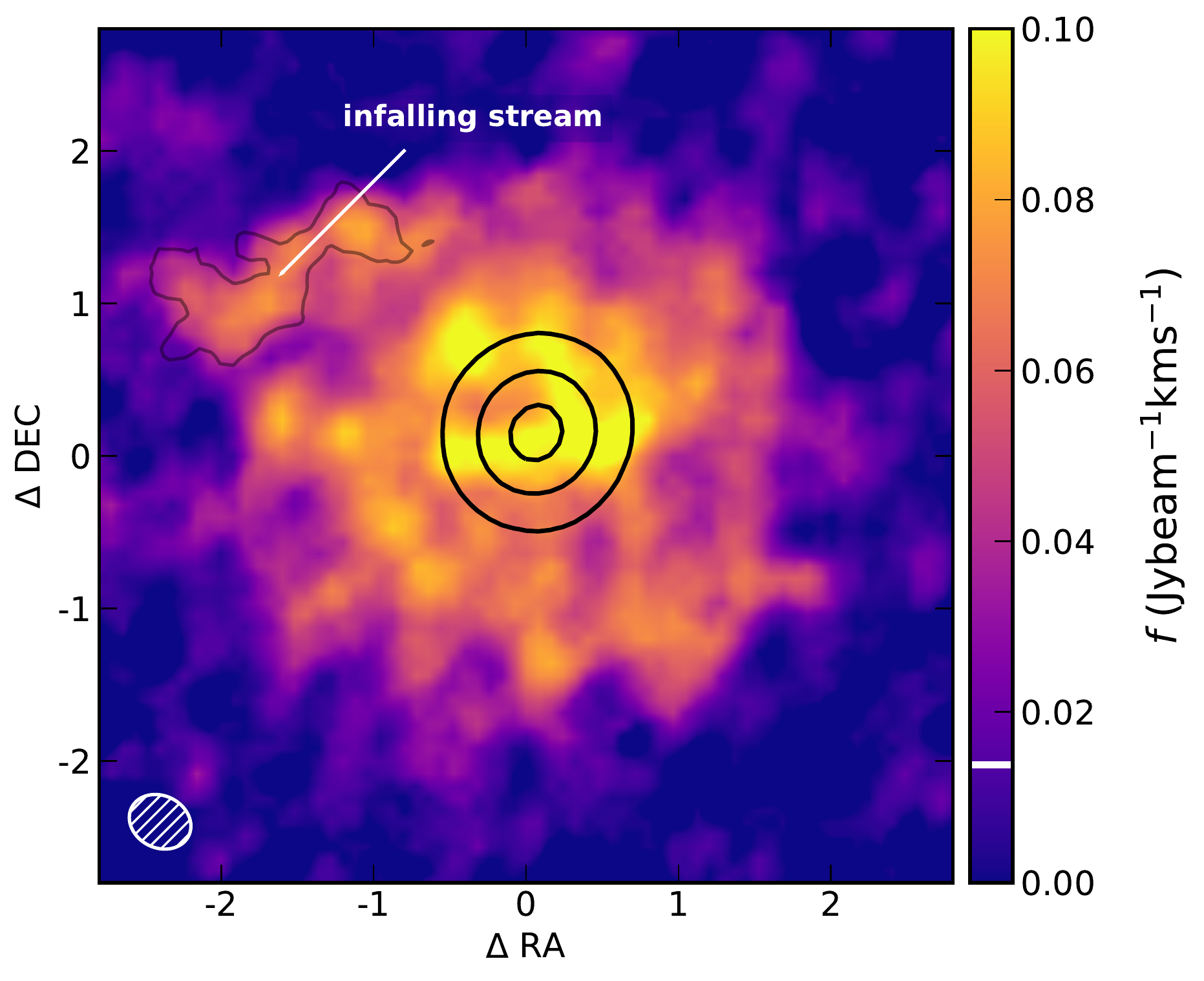}
    \caption{Total intensity or moment 0 map of the \cott $J$ = 3-2 emission in DR Tau.  The black contours show the continuum emission at 0.8 mm. The grey contours show the residuals after subtracting an azimuthally averaged radial disk profile to  highlight the infalling stream. The beam size is shown in the lower left corner.}
    \label{fig:dr_mom0co}
\end{figure}

The results of the line emission fit of DR Tau are presented in Fig. \ref{fig:dr_lines} for the best-fit model with a carbon depletion factor of 5.
A map of the surface brightness of the $^{13}$CO $J$ = 3-2 transition is presented in Fig. \ref{fig:dr_mom0co}.
Using least squares fitting, a Gaussian disk component with an additional Gaussian component centered at 0.7 km s$^{-1}$ accounting for the non-disk emission is fitted to the [C~I] and $^{13}$CO $J$ = 3-2 spectra to determine the disk contribution.

The low ratio between the [C~I] and CO isotopologue fluxes of the disk could only be explained by assuming $\gamma < 1$.
Decreasing $\gamma$ in the model results in a sharper exponential cut-off from the surface density, which affects the emitting region of the [C~I] emission more than the CO emitting layer.
Following the approach described in \citet{2018A&A...619A.113M}, we fitted the slope of the emission in the so-called pivot-region of the \cott and \coet emission where the line is optically thin, but has enough column density to efficiently self-shield against isotope selective processes.
We find $\gamma = 0.5$ and $\gamma = 0.8$ for the $^{13}$CO and C$^{18}$O radial intensity profiles, respectively. 
In the modeling we used $\gamma = 0.5$, which value is consistent with theoretical values for a viscously heated disk \citep{2009ApJ...705.1206C}.
%The model suggests that the gas disk is slightly more extended compared to the dust disk than expected from opacity arguments (see Appendix \ref{app:surface_density_DR}), which indicates that dust evolution and radial drift play an important role in the bulk of the outer disk.
%For the sake of simplicity, we do not include dust evolution in our modeling.

The C$^{18}$O fluxes together with [C~I] provide the strongest constraints on (C/H)$_{\rm gas}$. 
Therefore, we include the isotopologue-selective photodissociation of CO \citep{2014A&A...572A..96M,2016A&A...594A..85M}, which prevents underestimating (C/H)$_{\rm gas}$ or the gas mass.
We adopt ISM-like $^{13}$C and O$^{18}$ abundance ratios of 1.753 ppm and 0.5143 ppm with respect to H \citep{1999RPPh...62..143W}.
A model with (C/H) = 27 ppm, or a depletion factor of 5, reproduces the [C~I] emission and most CO isotopologue lines.
It overproduces the $^{13}$CO $J$ = 2-1 line by a factor of 3. 
The modeling suggests similar radial emitting regions for both CO transitions.
Using the peak ratio between the \cott and \coet $J$ = 2-1 spectra we determine the opacity of the line and the corresponding kinetic temperature using Eq. 1 and Eq. 2 from \citet{2016ApJ...823...91S}.
The peak ratio between the two CO isotopologue transitions is only 1.45, while a ratio of 8 is expected based on the ISM abundances of $^{13}$C and $^{18}$O. This results in an opacity of 9.4 and a temperature of the emitting layer of 42 K.
A similar analysis for the CO $J$ = 3-2 isotopologue transition results in a peak ratio of 2.85, an opacity of 3.3 and a kinetic temperature of the emitting layer of 30 K.
The reason why CO $J$ = 2-1 emission seems to be much more optically thick, even though it has a much lower Einstein $A$ coefficient, could be the lack of short spacing in the uv-plane which means that the data is not sensitive to large scale structures that are expected to be more dominant in $^{13}$CO than in \coet emission. In depth analysis of the temperature structure of the disk is beyond the scope of this paper.
We discuss the CO isotopologue emission and the methods used to constrain $\gamma$ in more detail in Appendix \ref{app:surface_density_DR}.

The best-fit carbon and oxygen abundances are found to be 2.7 x 10$^{-5}$ and 5.76 x 10$^{-5}$ which corresponds to a carbon and oxygen deficiency of a factor of 5$^{+2}_{-1}$.
The analysis of the photospheric abundance and the corresponding depletion factor found in the inner disk of DR Tau is $13^{+8}_{-4}$ \citep{2019A&A...632A..32M}. 
The low abundance in the inner disk in comparison with the outer disk is consistent with radial trapping of carbon-rich ices beyond the CO snowline.
We return to this in Sect. \ref{sec:discussion}.

The non-Keplerian, redshifted emission that is observed in both the [C~I] $J$ = 1-0 and \cott $J$ = 3-2 spectrum at a velocity of 0.7 km s$^{-1}$ originates likely from the circum-disk material that is also visible in the moment 0 map of the \cott emission to the northeast (indicated in Fig. \ref{fig:dr_mom0co}).
The high [C~I]/$^{13}$CO 3-2 flux ratio of this feature suggests that the emission is coming from material that is elevated above the disk surface and thus exposed to a high UV flux.
The contaminating material is confined to a single velocity, is not kinematically connected to the disk and may be linked with the late infalling material seen in scattered light observations \citep{2021arXiv211101702M}.
The C$^{18}$O 3-2 line is slightly broader than expected, likely caused by contamination with the same accreting material, but here we do not try to make a two-component fit.
Increasing the C/O ratio in a similar manner to our treatment of DO Tau in Sect. \ref{ssec:results_DO_Tau} could improve the fit to the $^{13}$CO $J$ 2-1 spectrum. However values of up to  3 are insufficient to fit the observed line flux. 
Furthermore, this would affect the higher $S/N$ $^{13}$CO $J$=3-2 emission as well, which is reasonably matched by models with the local interstellar standard C/O.
\begin{figure}[!t]
    \centering
    \includegraphics[width = 1\linewidth]{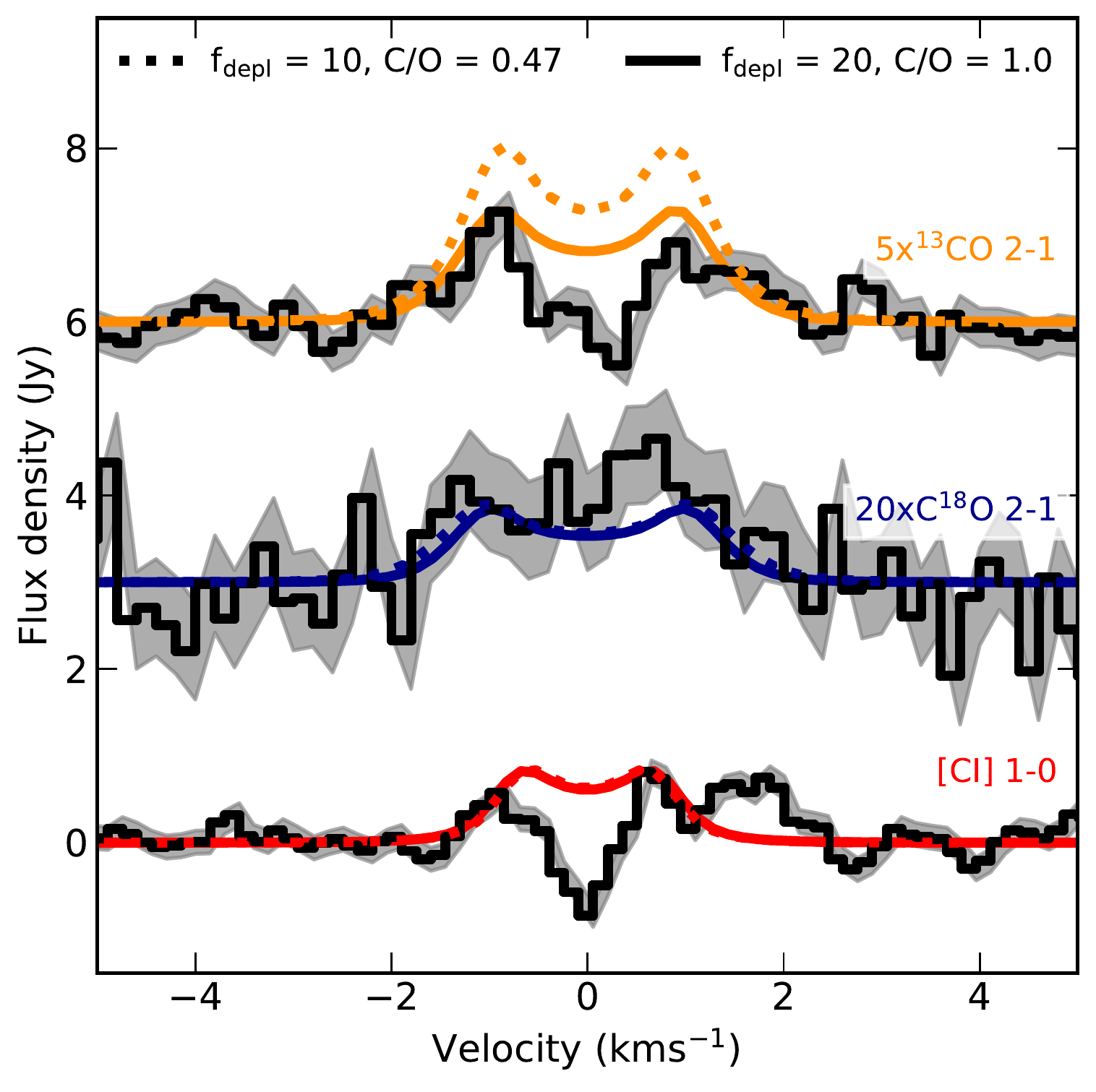}
    \caption{Observations and modeling of DO Tau. Two models are shown, the dashed line is a model with ISM C/O ratio of 0.47 and a carbon depletion factor of 10, the solid line is a model with an enhanced C/O ratio of 1 and a volatile carbon depletion fraction of 20. 1$\sigma$ errorbars are indicated in grey for each spectrum. The data and models of the \cott and \coet $J$=2-1 transitions have been multiplied by a factor of 5 and 20, respectively, for visualization purposes.}
    \label{fig:do_lines}
\end{figure}

\subsection{DO Tau: moderate depletion of carbon and oxygen}
\label{ssec:results_DO_Tau}
The data and best representative modeled spectra for DO Tau are illustrated in Fig. \ref{fig:do_lines}. 
We neglected the redshifted excess > 1 km~s$^{-1}$ which is associated with a pole-on outflow, as explained in the observation section and illustrated in Appendix \ref{app:diskwind_DO}.

The $^{13}$CO 2-1 line is optically thick, with a ratio of 2 with respect to the total \coet 2-1 flux, while a ratio of 8 is expected based on the relative abundance of the two isotopes in the diffuse ISM.
We present two models in Fig. \ref{fig:do_lines}, one with $f_{\rm depl}$ = 10 and an ISM-like C/O ratio and one with $f_{\rm depl}$ = 20 and an enhanced C/O ratio of 1.
In the later model, oxygen is less abundant with respect to carbon than the first model.
If oxygen is less abundant in the model, but the carbon abundance remains constant at ISM level, less CO reacts on the ice with OH to form CO$_2$ and H \citep[e.g.,][]{2011MNRAS.413.2281I}, thus the total CO abundance increases.
By using an increased C/O ratio there will be more carbon in atomic state, because there will be more CO in the atmospheric layers to be photodissociated in C$^{0}$ and C$^{+}$ and less oxygen to convert C back in CO. 
Since \cott is optically thick, changing the oxygen abundance does not change the total line flux of the model significantly. 
The best fitted C/O value based on the [C~I] and CO isotopologue emission is $\sim$1, which is about twice the ISM value.

C$_2$H can provide much stronger constraints on the C/O ratio than [C I] and CO, as it is predominantly formed in regions where C/O ratios are high \citep{2016ApJ...831..101B,2019A&A...631A..69M,2021ApJ...910....3B}.
The observed upper limit of C$_2$H = 73 mJy km s$^{-1}$ found by \citet{2019ApJ...876...25B} is low compared to the other sources in their sample, and more consistent with C/O close to the ISM value.
To prove this we ran an extended chemical model based on the chemical model from \citet{2018A&A...615A..75V} that includes C$_2$H and C$_2$H$_2$ chemistry using the same temperature and density structure as for the atomic carbon and CO emission.
We find that for a C/O ratio of 1 and carbon depleted by a factor of 20 with respect to the ISM value, the C$_2$H flux would be 550 mJy km s$^{-1}$.
An ISM-like C/O ratio of 0.47 and a carbon depletion fraction of 10 results in a total flux of 64 mJy km s$^{-1}$.
C$_2$H traces regions in the disks that are closer to the star than [C~I], which could mean that the discrepancy is caused by a radially varying C/O ratio in the disk. However, the systematic uncertainty of the models are too large to say anything conclusive about that, and full modeling of C$_2$H is beyond the scope of this paper.
Deeper C$_2$H observations are needed to better constrain the C/O ratio.
In the rest of this paper we use a C/H ratio that is 17$^{+3}_{-2}$ times lower than the ISM value, set by the carbon abundance grid in Fig. \ref{fig:chgrid}.

Our carbon abundance in the outer disk of DO Tau is similar to, or higher, than the abundance in the inner disk, which is found to be depleted with gas-phase carbon by a factor of 22$^{+20}_{-11}$ \citep{2019A&A...632A..32M}.
This radial pattern is consistent with retaining solids beyond the CO snowline.
We return to this in Sect. \ref{sec:discussion}.

\begin{figure}[!t]
    \centering
    \includegraphics[width = 1\linewidth]{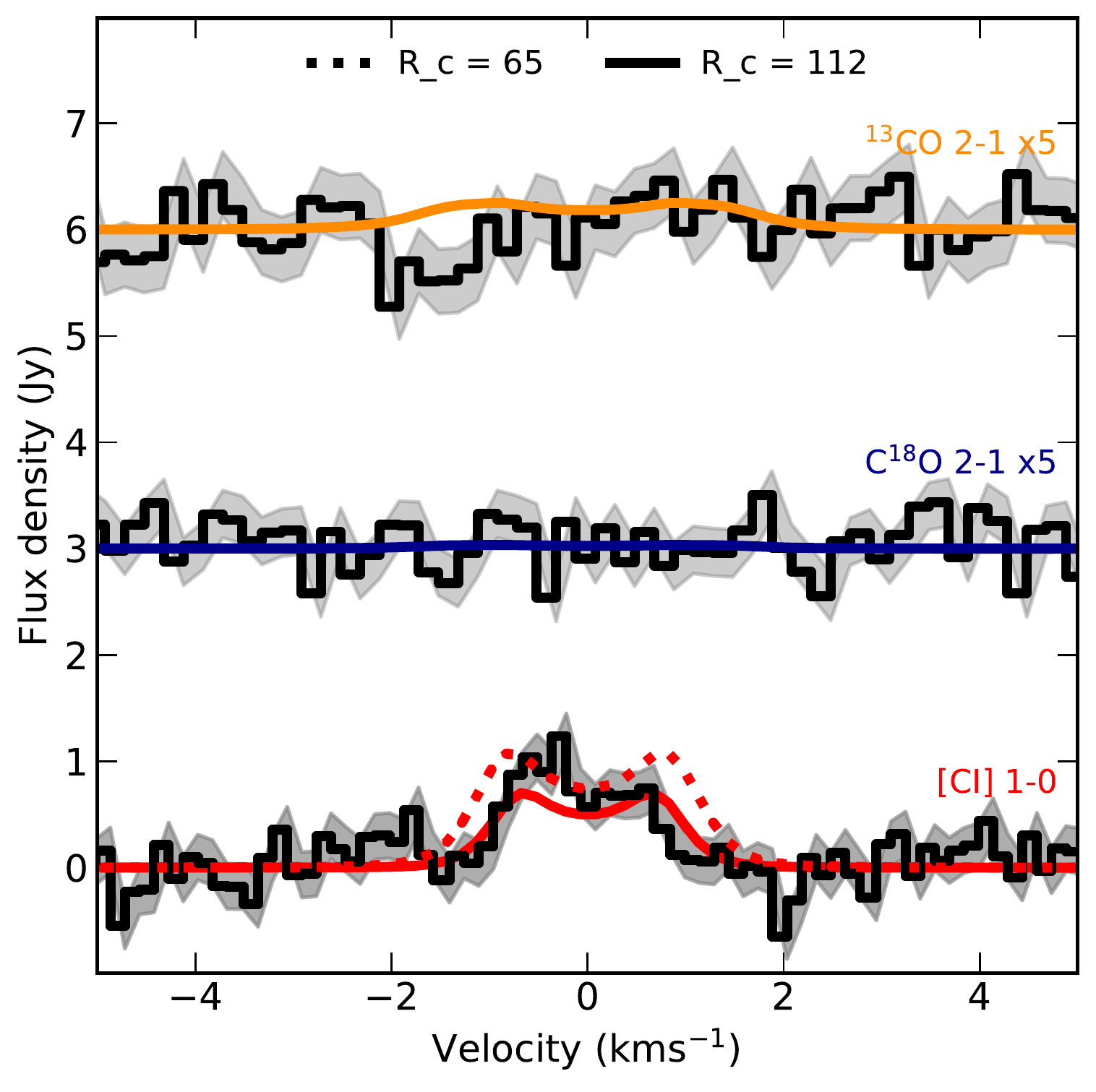}
    \caption{Observations and modeling of DL Tau. Two models are plotted, the solid line is a model with $R_\mathrm{c} = 112$ AU as observed in the dust, the dashed line as a model with $R_\mathrm{c}$ = 65, which fits the spatially resolved [C~I] emission best. Further details can be found in the main text. 1$\sigma$ errorbars are indicated in grey for each spectrum. The data and models of the weak CO isotopologue $J$=2-1 transitions have been multiplied by a factor of 5 for visualization purposes.}
    \label{fig:dl_lines}
\end{figure}

\subsection{DL Tau: severe depletion of carbon and oxygen}
The data and best representative modeled spectra for DL Tau are illustrated in Fig. \ref{fig:dl_lines}. 
The [C~I] spectrum has a disk component that consists of a clean double-peaked Keplerian profile, without cloud contamination. 
The two CO isotopologue $J$ = 2-1 lines are both upper limits, but the upper limits are low, so that they give additional constraints on the gas phase carbon abundance.

The best representative model for DL Tau has $R_{\rm c}$ = 112 AU, similar to the value found for the dust disk \citep{2015ApJ...808..102K,2019ApJ...882...49L}. 
However, convolving the modeled [C~I] emission with the same beam as the observations and comparing the azimuthally averaged radial profiles suggests that the radial extent is two times larger than the ACA beam of $\sim3\dprime$, while the observations appear as an unresolved point-source down to the same RMS level as the data.
The most likely explanation to this dichotomy is that the emission in the outer parts of the disk is buried in the noise.
To be consistent with the spatial information of the [C~I] emission, we find that $R_\mathrm{c}$ has to be smaller than 65 AU.
The SED constrains the total mass, but does not set limits to $R_{\rm c}$.
The characteristic radius of the dust is 112 AU  \citep{2015ApJ...808..102K,2019ApJ...882...49L}, and the CO outer disk radius is 600 $\pm$ 150 AU in SMA observations, which is consistent with $R_{\rm c}$ = 112 AU.
On top of that, a model with $R_{\rm c}$ = 112 AU fits the [C~I] $J$ = 1-0 line profile better than a model with the same mass, but with $R_{\rm c}$ = 65 AU (see Fig. \ref{fig:dl_lines}).
The line profile is, besides the disk mass distribution, also dependent on the mass of the star. 
There is some uncertainty on the masses of stars, based on their evolutionary track, but the stellar mass would have to be off by a factor of 2-3 to fit the Keplerian profile, which is beyond the usually assumed uncertainties for the stellar mass \citep{1998A&A...337..403B}.

For the R$_\mathrm{c}$ = 112 AU model, we find a best fitting depletion factor of 157$^{+17}_{-15}$. 
If the disk were smaller, with $R_\mathrm{c}$ = 65, DL Tau would still be depleted in carbon by a factor of 52$^{+6}_{-8}$.
The high depletion value for DL Tau is consistent with the age of the source, as shown by chemical modeling of CO emission \citep{2020ApJ...899..134K}.
We return to this in Sect. \ref{ssec:comparison_modeling}.
In order to put further constraints to the model for DL Tau and exclude that the disk is depleted in gas, resulting in a small outer gas radius, we need ALMA $^{12}$CO and CO isotopologue data and high resolution [C~I] observations with more sensitivity in the outer parts of the disk.

\subsection{Systematic uncertainty}
Additional models are presented for the three sources in which the most relevant disk parameters are varied in order to obtain an indication of the systematic uncertainty on the carbon abundance in the modeling, as discussed in their respective subsection.
These additional models show that the carbon depletion factor can be trusted within a factor of 3, with the assumption that the gas-to-dust ratio is equal to the canonical value of 100.
\citet{2016A&A...588A.108K} showed how the C/H abundance systematically depends on a large number of other disk parameters, including the dust-to-gas ratio. They found that in the worst case scenario, when taking into account a possibly varying dust-to-gas ratio, the [C~I] flux in a disk is robustly sensitive to carbon depletions of at least a factor of 10 in the outer disk. Therefore, we can conclude that the significant carbon depletions in DO Tau and DL Tau are robust to the systematic uncertainties. In contrast, the order of magnitude carbon depletion in DR Tau, while significant under our current model assumption of a gas-to-dust ratio of 100, could be consistent with the ISM carbon abundance if the full range of systematic uncertainties in the DALI models is taken into account.

\section{Discussion}
\label{sec:discussion}
We present the first successful ALMA mini-survey of neutral carbon to constrain the elemental carbon depletion in the outer disk of a sample of seven T Tauri disks.
Two sources in our sample, DR Tau and DO Tau, have moderate depletion of volatile carbon and oxygen in the outer disk, and one source, DL Tau, has severe depletion.
The observed sources show two distinct radial carbon abundance behaviors: the moderately depleted sources have carbon abundances in the outer disk that are consistent with the inner disk, while the severely depleted source has a much higher carbon abundance in the inner disk than in the outer disk.
This work strengthens the case that some sources are depleted in gas-phase carbon in the outer disk in addition to direct freeze-out and photodissociation of CO.
Fig. \ref{fig:depletion_categories} summarizes our results of the carbon abundance in the outer disk compared to the inner disk.
We added TW Hya \citep{2019ApJ...883...98Z,2016A&A...592A..83K,2020A&A...642L..15M}, HD 163296 \citep{2015A&A...582L..10K,2019ApJ...883...98Z} and HD 100546 \citep{2015A&A...582L..10K,2016A&A...592A..83K} and a combined point for ten Class 0/I sources \citep{2014A&A...562A..77H,2020ApJ...891L..17Z,2020ApJ...898...97B} to the figure for reference. 
These three sources are the only other sources for which the carbon abundance is determined in both the inner disk and outer disk.
For the Herbig Ae/Be stars we make the assumption that the composition of the stellar photo-sphere is representative of the accreting material in the inner disk, as in \citet{2015A&A...582L..10K}.

\subsection{Implications of the radial abundance profile}
\begin{figure}[!t]
    \centering
    \includegraphics[width = \linewidth]{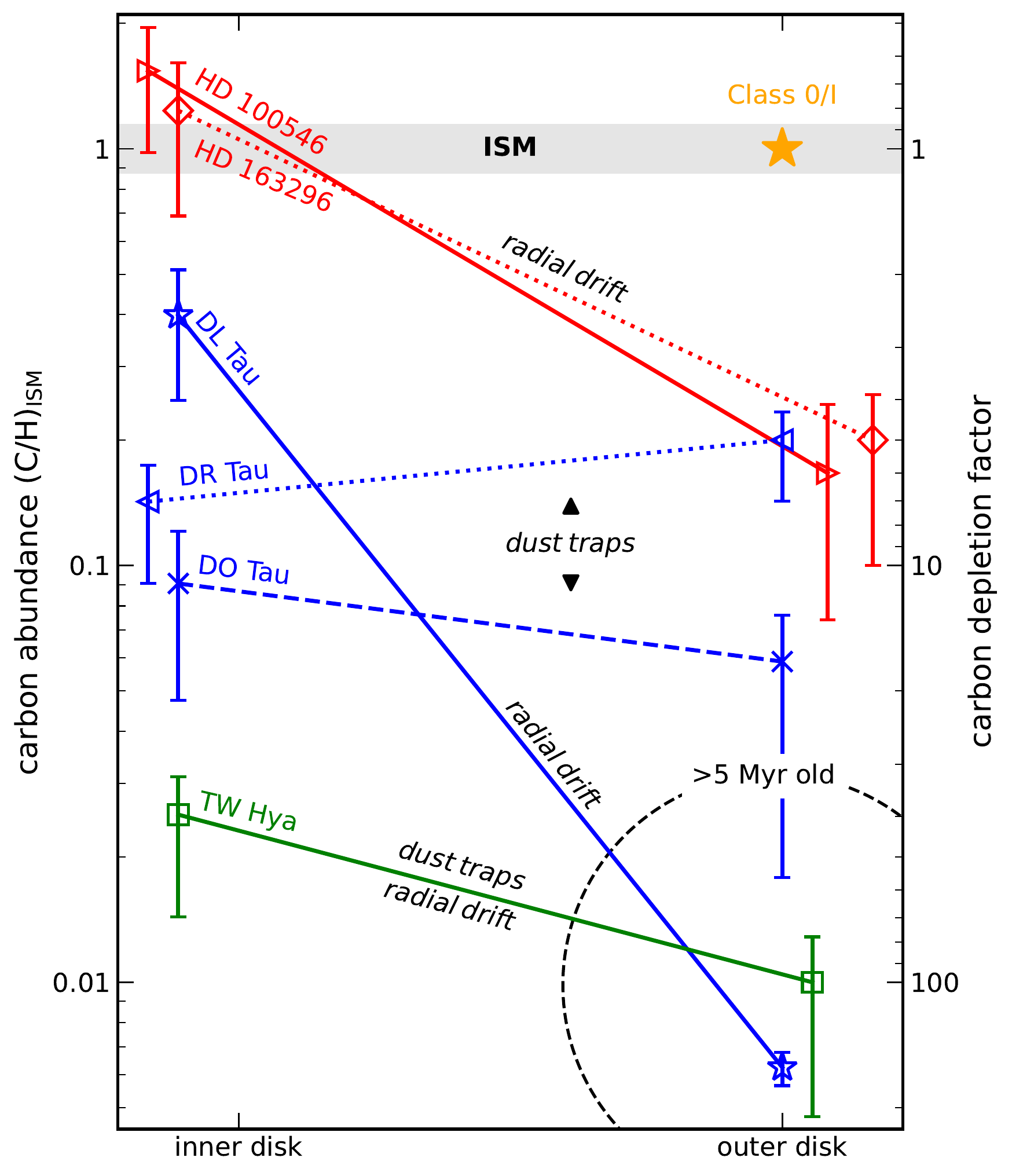}
    \caption{Radial profile of the volatile carbon depletion fraction based on one point inside the dust sublimation rim and a disk averaged carbon abundance in the outer disk. The outer depletion fractions of the sources in blue are from this work. Inner disk carbon abundances of these three sources are taken from \citet{2019A&A...632A..32M}. We added TW Hya in green \citep{2019ApJ...883...98Z,2016A&A...592A..83K,2020A&A...642L..15M} and one point in the outer disk for ten Class 0/I sources in yellow \citep{2014A&A...562A..77H,2020ApJ...891L..17Z,2020ApJ...898...97B} to illustrate the increase in the level of carbon depletion during the system's evolution.
    Two herbig Ae/Be stars, HD 163296 \citep{2015A&A...582L..10K,2019ApJ...883...98Z} and HD 100546 \citep{2015A&A...582L..10K,2016A&A...592A..83K} are shown in red to complete the picture of carbon depletion that we have. These sources do not follow the evolutionary trend, because they are much warmer.}
    \label{fig:depletion_categories}
\end{figure}

Almost all sources with observed elemental carbon abundances have a considerable increase in carbon abundance in the inner disk with respect to the outer disk, except for the two young and compact disks in our sample, DR Tau and DO Tau (see Fig. \ref{fig:depletion_categories}).
Large dust grains, the formation of which causes gaseous carbon depletion, will eventually drift inwards towards the star, driven by the headwind of the gas that moves slightly sub-Keplerian due to the pressure support.
With efficient drifting of dust grains and without any dust trapping, the carbon will be released inside the CO snowline. 
Any carbon that is processed into less volatile or refractory structures will ultimately sublimate at the dust sublimation rim.
The inner disk gas (inside the dust sublimation rim) is constantly replenished with carbon-poor gas from the outer disk, which means that an enhancement there has to be a result of the sublimation of carbon-rich ices as a result of radial drift.

\subsubsection{Dust traps}
\label{ssec:dust_traps}
Strong pressure maxima and efficient planetesimal formation locations, incorporating the grains in larger bodies, can both stop the grains from radial drifting.
DL Tau has three rings in the dust continuum observations that are located outside the CO snowline \citep{2019ApJ...882...49L}. 
Sufficiently large dust grains will be trapped in the pressure maxima at the edge of these gaps, so that those grains remain in the outer disk, but trapped very locally, which means that the CO ice that has formed on these grains cannot efficiently return to the gas phase.
The dust trapping in the pressure maxima that are observed in DL Tau are efficient enough to prevent a super-solar carbon abundance in the inner disk contrary to what is observed in the radial CO abundance profiles of other protoplanetary disks like HD 163296 \citep{2019ApJ...883...98Z,2020ApJ...891L..16Z}.
However, the fact that we see an enhancement in the inner disk with respect to the outer disk implies that either the traps are ``leaky", i.e., not 100\% efficient at trapping dust, or that they formed in a later stage of the disk evolution, i.e., shorter time ago than the viscous time needed to accrete the material from the dust traps to the inner disk, or that carbon depletion mechanisms started later than the viscous time.
The last option is unlikely as recent studies pointed out that volatile CO rapidly evolves during the Class I and early Class II stage \citep{2020ApJ...891L..17Z,2020ApJ...898...97B}. 

The formation of a perfectly efficient dust trap at the CO snowline would prevent depleted carbon in the form of CO ice from entering the gas interior to the CO snowline. 
However, before it is observable via the near-infrared lines used by \citet{2019A&A...632A..32M} this depleted carbon would still need to viscously accrete from the CO snowline to $\sim$0.1 AU on a timescale of:
\begin{equation}
    t_{visc} = \frac{r^2\Omega_\mathrm{K}}{\alpha c_\mathrm{s}^2}
\end{equation}
with $\Omega_\mathrm{K}$ the Keplerian angular velocity and $c_\mathrm{s}$ the sound speed in the midplane of the disk given by $\sqrt{k_\mathrm{B}T/(2.3m_\mathrm{p})}$ with $k_\mathrm{B}$ the Boltzmann constant, $T$ the midplane temperature, $m_\mathrm{p}$ the proton mass, and $\alpha$ the viscosity parameter \citep{1973A&A....24..337S}.
The innermost gap outside of the CO snowline in DL Tau is located at $\sim$40 AU, and the temperature at the midplane is provided by \texttt{DALI}. 
However, $\alpha$ is not determined directly for DL Tau and is subject to large uncertainty. 
Observational measurements of $\alpha$ from turbulent broadening of CO $J$=2-1 emission lines range from $\alpha$ = 0.08 in DM Tau to $\alpha$ < 0.006 in MWC 480, while theory predicts still lower $\alpha$ around 10$^{-4}$ \citep[][and references therein]{2020ApJ...895..109F}.
Using $\alpha = 0.08$ and $\alpha$ = 0.006, we find that the formation of the dust trap at the CO snowline should be less than 0.16-2.2 Myr ago.
This could not explain DL Tau's carbon depleted inner disk if the substructure were formed at the start of the Class II phase ($\sim$7 Myrs ago), as \citet{2020ApJ...899..134K} showed that for a dust trap with >99\% trapping efficiency, the inner disk would be depleted within 3 Myr by more than the factor three with respect to the ISM abundance that is observed in DL Tau.
However, if $\alpha$ were < 0.003, which still agrees well with the observations, the viscous timescale would be long enough to cover even substructure that has formed right at the start of the Class II phase.

An alternative explanation for the large discrepancy between the carbon abundance in the inner and outer disk is that the dust traps in DL Tau are not 100\% efficient in trapping dust.
This idea is supported by modeling that shows that low drift efficiency will result in less enhancement in the inner disk \citep{2020ApJ...899..134K}.
\citet{2020A&A...642L..15M} devised a simple analytic model to calculate dust retention efficiencies.
The total dust trap efficiency at different positions in the TW Hya disk, based on the distribution of carbon is 99.8\%, hence the strong depletion in the inner disk (Fig. \ref{fig:depletion_categories}). 
Following equations 1-3 in Appendix C of \citet{2020A&A...642L..15M}, we derive a simple expression for the dust trapping efficiency of our disks beyond the inner dust sublimation radius:
\begin{equation}
    \label{eq:fl}
    f_\mathrm{trap} = 1 - \frac{X_{\rm g,in}-X_{\rm g,out}}{X_{\rm g,ISM}-X_{\rm g,out}},
\end{equation}
where $X_{\rm g,in}$ is the carbon abundance in the inner disk, $X_{\rm g,out}$ the carbon abundance in the outer disk and $X_{\rm g,ISM}$ the carbon abundance in the ISM.
Using this formula we find that the dust trap efficiency of the DL Tau disk is 61\%, assuming an inner disk abundance that is lower by a factor of 2.5 with respect to the ISM value \citep{2019A&A...632A..32M}.
Note that this approximation is highly susceptible to variation in carbon abundance close to the solar value, hence an upper limit. 
If no carbon depletion is assumed in the inner disk, which is still within the systematic uncertainty of the measurement, then the locking fraction would be 0\%.

\citet{2020ApJ...898...36L} find low spectral index, $\alpha$ = 2-2.5, in large regions of the disk of DL Tau by comparing resolved 1.3 mm and 2.9 mm ALMA continuum data.
They find similar disk outer radii at both wavelengths and show that this is consistent with a dust evolution model that includes a long lasting pressure bump.
A low spectral index is in line with advanced grain evolution, which is also expected based on the evolutionary stage of the source.
These authors show also that the spectral index in DL Tau varies only by $\sim$10\% near the pressure maxima, which is in line with the low trapping efficiencies of the dust traps we find in this work.
Note, however, that this low variation is a lower limit, since the gaps are unresolved and therefore less pronounced due to beam-smearing.

\subsubsection{Dust retention in smooth, compact systems}
The carbon abundances found for DR Tau and DO Tau are consistent within the observational uncertainties with what is observed in the inner disk, based on the accreting material onto the star \citep{2019A&A...632A..32M}.
This implies either that the carbon in ices on dust grains is prevented from drifting towards the star or that the chemistry outpaces the pebble drift by converting CO to less volatile species such as CO$_2$ and CH$_3$OH, preventing a resurgence of gas-phase CO just inside the snowline \citep{2020ApJ...899..134K}.
Based on the approximation of the locking fraction in Eq. \ref{eq:fl}, both DR Tau and DO Tau have a trapping efficiency close to 100\% (97\%, and >100\% for DO Tau and DR Tau respectively).
Both freeze-out and dust evolution scale with the orbital time scale, which means that carbon is more efficiently depleted closer in to the star, but outside the CO snowline. 
In extreme scenarios, this could result in a lower abundance in the inner disk than in the disk averaged abundance if radial drift is stopped close to the CO snowline and the inner disk is accreting the gas that is locally highly depleted in carbon just outside the CO snowline.
Also, additional dust processing could happen in the 20-40 AU between the CO snowline and the inner disk that could influence the carbon abundance.
This could be tested by measuring the C/H ratio with JWST or CRIRES+ data in the future.
Better resolved radial carbon abundance profiles are necessary to put further constraints on the amount of material that is left behind in the disk to form planets.

There are no signs of dust trapping pressure maxima in the dust continuum images of both DO Tau and DR Tau, often assumed to be a result from planets \citep{2019ApJ...882...49L}.
However, small substructure will be unresolved in the $\sim$15 AU beam of the continuum observations and the dust is optically thick up to $\sim$25 AU, beyond the CO snowline, which could hide the presence of a gap or dust ring.
Future high resolution studies, revolving around substructure near the habitable zone of compact disks, are necessary to determine whether substructure is exclusively present in extended systems or that compact disks host more substructure than we think.
For DR Tau, R$_\mathrm{gas}$/R$_\mathrm{dust}$ = 3.5 based on the 95\% effective radius of the \cott $J$ 3-2 emission and millimeter continuum, which is close to be a clear sign of dust evolution and radial drift \citep[see][]{2019A&A...629A..79T}.
Considering that the method in \citet[][]{2019A&A...629A..79T} is based on $^{12}$CO observations, which in general has a more extended emitting area than the \cott for opacity arguments, and the fact that the spatial extent of the CO isotopologue data is slightly larger than can be explained by the modeling (see Appendix \ref{app:surface_density_DR}), these systems likely have advanced dust grain evolution and radial drift.
This contrast with the radial carbon abundance profile could potentially be explained by dust traps relatively close to the star, not affecting the radial drift in the outer disk.
It is also possible that these systems have had efficient radial drift in the past, which is stopped due to recent formation of one or more planetary cores, but not enough to show up in the sub-millimeter continuum.

\subsection{Evolutionary trend in carbon depletion}
\label{ssec:comparison_modeling}
Combining all sources of which the carbon abundance in the outer disk is determined using multiple carbon carrying molecules (Fig. \ref{fig:depletion_categories}), we find a potential evolutionary trend in the level of elemental gas-phase carbon depletion, similar to those found in CO \citep{2020ApJ...891L..17Z,2020ApJ...898...97B}.
Recent measurements of C$^{18}$O in ten Class 0/I sources in multiple star forming regions \citep[shown as a yellow star in Fig. \ref{fig:depletion_categories}][]{2014A&A...562A..77H,2020ApJ...891L..17Z,2020ApJ...898...97B} indicate that the youngest sources are not yet depleted in CO, relative to the ISM value, therefore likely not in elemental carbon.
As disks age the snowline moves inwards and the vertical snow surface decreases, leading to the presence of a massive cold trap in the outer disk. 
Any CO that mixes through the dust growth layer or midplane will be lost to never return to the gas.
This leads to gas-phase carbon depletion as seen in the outer disks of DO Tau and DR Tau by 1-2 Myrs (blue crosses and triangles, respectively in Fig. \ref{fig:depletion_categories}). 
As the disks cool further, and CO-rich icy pebbles drift inward over time, the outer disk becomes increasingly depleted, leading to depletion by up to 2 orders of magnitude within 8 Myrs as in DL Tau and TW Hya (blue stars and green squares, respectively in Fig. \ref{fig:depletion_categories}). 
DL Tau has a similar age and size as TW Hya, and has also an extended ring system, which explains the striking similarity between the two sources in terms of their outer disk carbon depletion. 
The only dissimilarity arises from the fact that DL Tau is less efficient in trapping dust, resulting in a much larger difference between the inner and outer disk carbon abundances, which we addressed in Sect. \ref{ssec:dust_traps}.
TW Hya has an inner cavity \citep{2016ApJ...820L..40A}, which might act as an inner pressure bump that prevents the final drift of pebbles and reduce the carbon enhancement in the inner disk.
Similarly old disks around hotter stars, e.g. $\sim$ 10 Myr Herbig Ae/Be stars HD 163296 and HD 100546 (red diamonds and triangles, respectively in Fig. \ref{fig:depletion_categories}), remain only as depleted as the younger T Tauri disks, possibly due to increased heating and UV radiation from the central star preventing efficient CO ice formation over as large an area as for the T Tauri stars.  
Larger samples of disks are necessary in order to validate this evolutionary trend.

Calculations of the stellar age can vary considerably depending on the methods used to determine the effective temperature and luminosity to place the sources on evolutionary tracks. 
The ages for the sources in Taurus that we use here were calculated by \citet{2019A&A...632A..32M} from the \citet{2000A&A...358..593S} evolutionary tracks using stellar properties determined self-consistently from near-infrared spectra. The age estimates for DR Tau and DO Tau agree well with the young age of 1-2 Myr of the Taurus-Aurigae star forming region \citep{1995ApJS..101..117K}, and are close to literature values, 0.37-2.8 Myr and 0.37-0.93 Myr, respectively \citep[see e.g.,][]{2013ApJ...771..129A}. 
The age of DL Tau is relatively high for the Taurus-Aurigae star forming region and compared to the literature values of 0.7-2.6 Myr \citep{2013ApJ...771..129A} and 3.5$^{+2.8}_{-1.6}$ Myr \citep{2019ApJ...882...49L}. 
These differences are a result of \citet{2013ApJ...771..129A} and \citet{2019ApJ...882...49L} not taking into account the high veiling of DL Tau \citep[$r_\mathrm{J}\sim$2;][]{2019A&A...632A..32M} when determining the luminosity, namely using a direct fit to the dereddened photometry, that results in overestimating of the stellar luminosity by a factor of 2 and therefore underestimating of the stellar age. 
Knowing the precise stellar properties is therefore crucial in our understanding of the evolutionary sequence for the outer regions of the disks.

Recent dynamical modeling of CO depletion processes shows that only a combination of chemical processing, freeze-out and sublimation, vapor diffusion, dust and ice dynamics, and pebble formation and dynamics can explain the level of cold, gaseous CO depletion observed in CO surveys \citep{2020ApJ...899..134K,2018ApJ...856...85S}.
The processes that are involved have in general long timescales, requiring several Myrs to achieve appreciable levels of carbon depletion in the outer regions of a disk.
\citet{2020ApJ...899..134K} find that multiple orders of volatile carbon depletion can only be explained by efficient depletion integrated over at least 3 Myr. 
Our observations are in line with this modeling, and show that the high level of outer disk carbon depletion of DL Tau and TW Hya is likely due to the fact that more volatile CO has had the chance to freeze out and be locked up in more complex molecules, in large dust grains or transported away from the outer disk by radial drift.

A secondary impacting mechanism on the level of depletion may be the size of the disks, either set by evolutionary or mass arguments. 
The carbon depletion mechanism is thought to be more efficient in disks with a larger freeze-out zone.
HD 100546 and HD 163296 both have no strong carbon depletion, even though the properties of the disk are similar to TW Hya and both sources may be even older than TW Hya \citep{2016A&A...592A..83K}.
However, these two sources are much warmer because of the bright star in the centre, hence they have a small region where CO can freeze out.
Similarly, DO Tau and DR Tau are both compact, thus have a relatively small volume beyond the CO snowline where CO freeze-out may occur.
Future observations and modeling are needed to get a better understanding of the carbon depletion processes during the different stages of evolution of the disk.

\section{Conclusions}
\label{sec:conclusions}
In this paper, we present observations and modeling of the main gas-phase carbon carriers in protoplanetary disk atmospheres, C$^{0}$ and CO isotopologues. We observed a sample of seven disks around T Tauri stars in [C~I]. Six out of the seven sources have a firm detection in [C~I], in four of the spectra we can constrain the disk component.
Three of the sources are modeled with the thermo-chemical disk model \texttt{DALI} to determine the C/H ratio in the outer disk.
We can conclude the following
\begin{itemize}
    \item We find depletion factors in gas-phase elemental carbon abundance of 5$^{+2}_{-1}$, 17$^{+39}_{-7}$ and 157$^{+17}_{-15}$ for the outer disks of DR Tau, DO Tau and DL Tau, respectively.
    \item DL Tau is much more depleted in carbon in the outer disk than in the inner disk, following a similar pattern as observed in other disks. This is likely the result of advanced grain evolution, efficient radial drift and little locking of carbon in the outer disk. We propose that either the substructure formed later than the viscous time, or that the dust rings observed in the millimeter continuum images of this disk are leaky, i.e., have <100\% efficiency in trapping dust.
    \item DO Tau and DR Tau show, within the systematic error, similar depletion in volatile carbon in the outer disk as in the inner disk that indicates very efficient dust trapping in the outer disk.
    Though both disks do not show any dust substructures from current high resolution millimeter continuum observations, our radial carbon abundance profiles suggest the presence of small scale pressure maxima in the inner disk of these compact dust disks. In addition, these systems could have had efficient planetesimal formation that stopped most of the radially drifting dust grains. 
    \item Combining all literature volatile carbon depletion factors in inner and outer disks of T Tauri stars, we find hints for an evolutionary trend in carbon depletion that is consistent with CO isotopologue studies and dynamical models of CO depletion processes that show increasing depletion for older sources.
\end{itemize}
Expanding the number of points in the radial profiles of carbon depletion will be crucial to understand where the dust is locked up in the disk, potentially in the form of small planetesimals. Future higher spatial resolution observations of the [C~I] $J$ = 1-0 line, in combination with high resolution CO and continuum data, have the potential to be one of the few ways to probe directly the formation of small planetesimals in T Tauri disks.

\begin{acknowledgements}
This paper makes use of the following ALMA data: \\
ADS/JAO.ALMA\#2017.1.00857.S,  \\
ADS/JAO.ALMA\#2016.1.00158.S, \\
ADS/JAO.ALMA\#2016.1.01164.S, \\
ALMA is a partner-ship of ESO (representing its member states), NSF (USA), and NINS (Japan), together with NRC (Canada) and NSC and ASIAA (Taiwan), in cooperation with the Republic of Chile. The Joint ALMA Observatory is operated by ESO, AUI/NRAO, and NAOJ. We thank the referee for comments that helped to
improve the manuscript. 
\end{acknowledgements}

\bibliographystyle{aa}
\bibliography{references.bib}

\begin{appendix}
\section{Observing details}
In Table \ref{tab:observing_log} we present the observing log of the [C~I] observations.
    \begin{table*}[!t]
    \caption{ALMA Observing Log}
    \label{tab:observing_log} 
    \begin{tabular}{L{3.5cm}L{1.5cm}|C{2.5cm}|C{2cm}C{3cm}C{2cm}}
    \hline
    \hline
    \Tstrut Execution blocks & N$_\mathrm{ant}$ \Tstrut& Calibrators \Tstrut& Targets \Tstrut& \Tstrut Integration time & Beam \Tstrut\\
    (UTC Time) & &&&(s)& ($\dprime$)\\ 
    (1)& (2)& (3)& (4)& (5) & (6)\\
    \hline
    2018 Aug 17\Tstrut& 9 \Tstrut&J0440+1437\Tstrut&DL Tau \Tstrut&5080\Tstrut& 3.6 x 2.8\Tstrut\\
    2018 Aug 28& 11&J0510+1800&DO Tau&5291&3.5 x 2.9\\
    2018 Sep 3& 11&J0522-3627&DR Tau&3387&3.4 x 2.7\\
    2018 Sep 4&10&&FZ Tau&4656&3.6 x 2.7\\
    2018 Sep 5&10&&&\\
    2018 Sep 11&12&&&\\
    2018 Sep 12&11&&&\\
    \hline
    2018 Aug 25\Tstrut&10\Tstrut&J0522-3627\Tstrut&FM Cha \Tstrut&8376\Tstrut&3.7 x 3.1\Tstrut\\
    2018 Sep 18&12&J1058-8003&WW Cha &8528&3.2 x 2.9\\
    2018 Sep 19&11&J1145-6954&\\
    2018 Sep 19&11&\\
    2018 Sep 20&11&\\
    2018 Sep 22&&\\
    \hline
    2018 Sep 9\Tstrut&\Tstrut10& J1256-0547\Tstrut &AS 205 A \Tstrut&2238\Tstrut& 3.1 x 2.1\Tstrut\\
    2018 Sep 10&11& J1553-2422\\
    && J1517-2422\\
    \hline\\
    \end{tabular}\\
    \end{table*}

\section{pole-on outflow in [C~I]}\label{app:diskwind_DO}
\begin{figure*}[!b]
    \includegraphics[width = 0.95\linewidth]{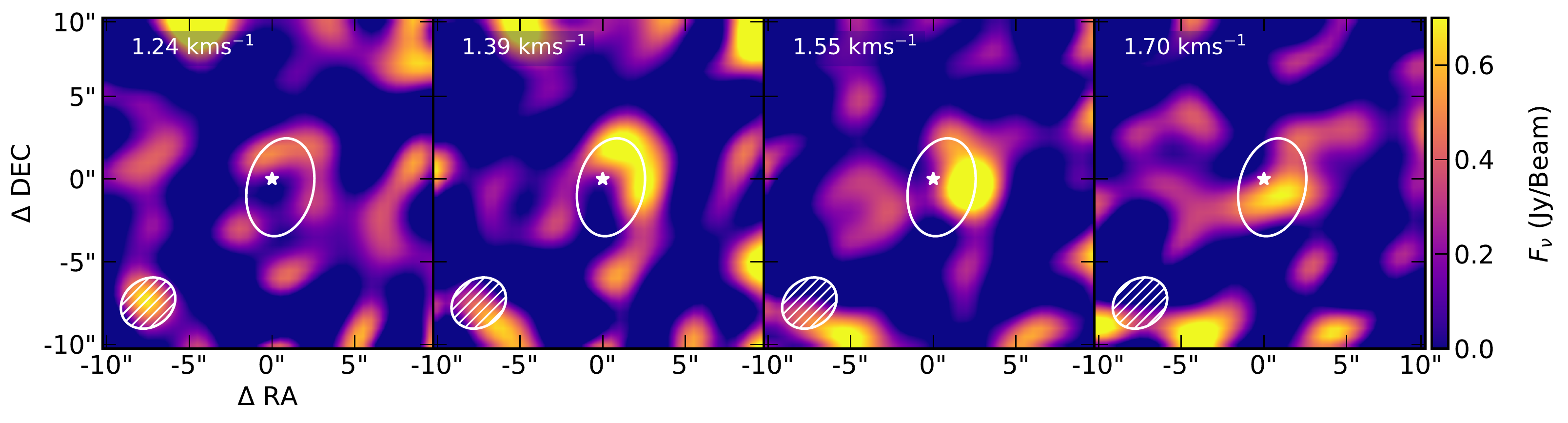}
    \caption{Channelmaps of the [C~I] $J$ = 1-0 emission of the non-Keplerian redshifted component in DO Tau. Channel velocities are relative to the cloud velocity. The white ellipse denote the best-fit ellipse to the pole-on outflow observed in CO ($B_{\rm maj}$ = 3$\dprime$, $B_{\rm min}$ = 2$\dprime$, PA = 167$^{\rm o}$ \citep{2020AJ....159..171F}). The arc-like emission shares striking similarity with the observed pole-on outflow observed in CO.}
    \label{fig:do_diskwind}
    \end{figure*}

The redshifted excess emission in the spectrum of DO Tau (see Fig. \ref{fig:spectra}) originates from an off-center, spatially resolved ring of emission, likely emitted from the pole-on outflow that is also observed in CO \citep{2020AJ....159..171F}. 
Some of the channels at this velocity are presented in Fig. \ref{fig:do_diskwind}.
The white ellipse in this figure marks the best-fit ellipse to the pole-on outflow observed in $^{12}$CO ($B_{\rm maj}$ = 3$\dprime$, $B_{\rm min}$ = 2$\dprime$, PA = 167$^{\rm o}$) by \citet{2020AJ....159..171F}.
Studying the outflow in detail is beyond the scope of this paper; additional research to the nature of the wind and its impact on the disk dynamics is necessary.
    
\section{Surface density profile DR Tau}\label{app:surface_density_DR}
\begin{figure*}[t]
    \centering
    \includegraphics[width = .9\linewidth]{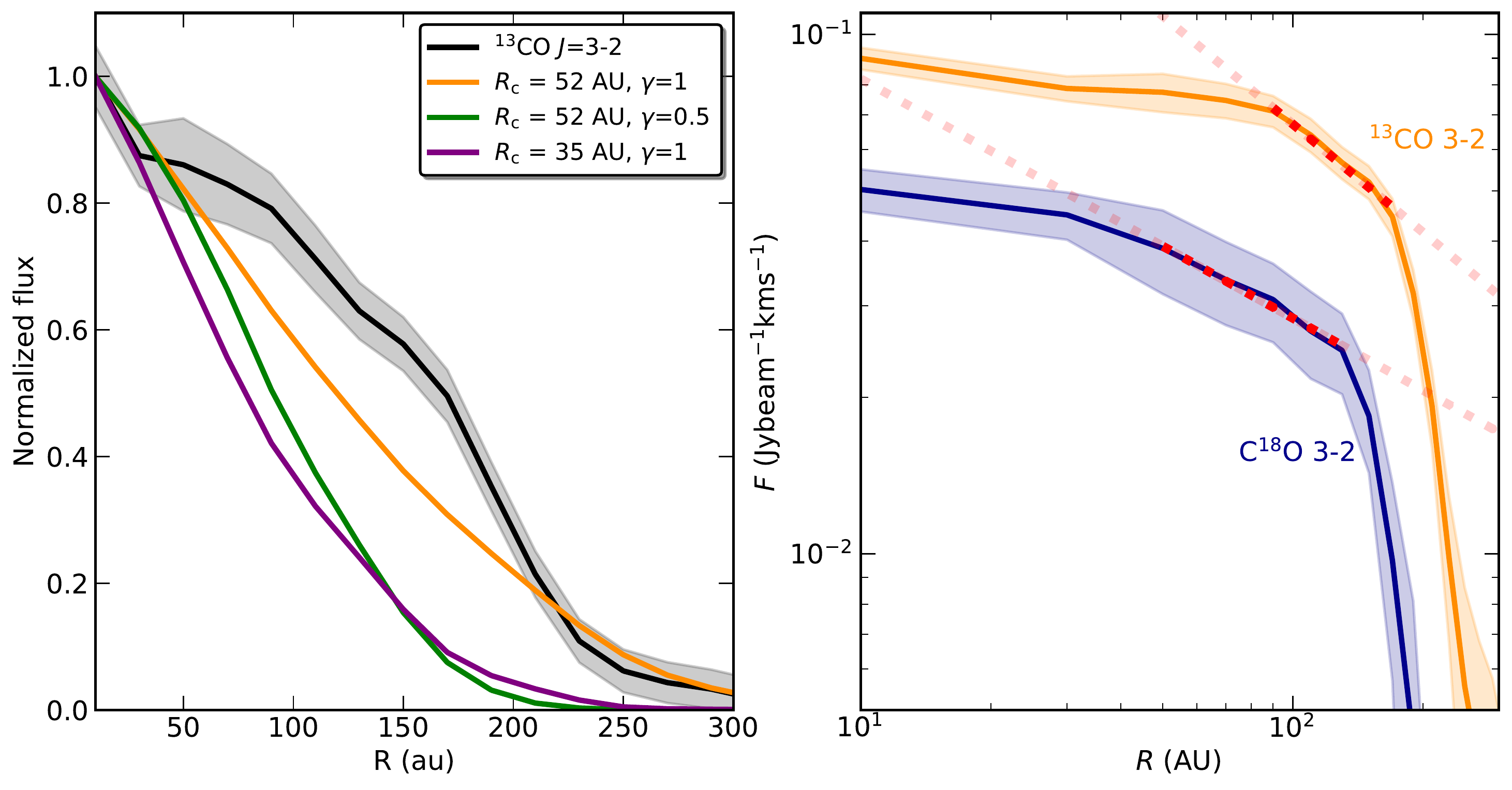}
    \caption{Left: Radial profile of the $^{13}$CO data in black with three models varying $R_\mathrm{c}$ and $\gamma$ convolved with a gaussian beam, similar to the observations. Parameters that are changed with respect to the fiducial model used in the main text are denoted in the legend.
    The shaded region marks the 1$\sigma$ uncertainty on the azimuthal average of the data. Right: Radial profile of the $^{13}$CO (orange) and C$^{18}$O emission (blue) in DR Tau. The dotted lines are a linear fit in log-log space. The fit is only performed in the pivot-region (bold dotted line) where the line is optically thin, but with enough column density to efficiently self-shield. The shaded regions mark the 1$\sigma$ uncertainty on the azimuthal average of the data.}
    \label{fig:dr_profile}
\end{figure*}

\begin{figure}[!t]
    \centering
    \includegraphics[width = .95\linewidth]{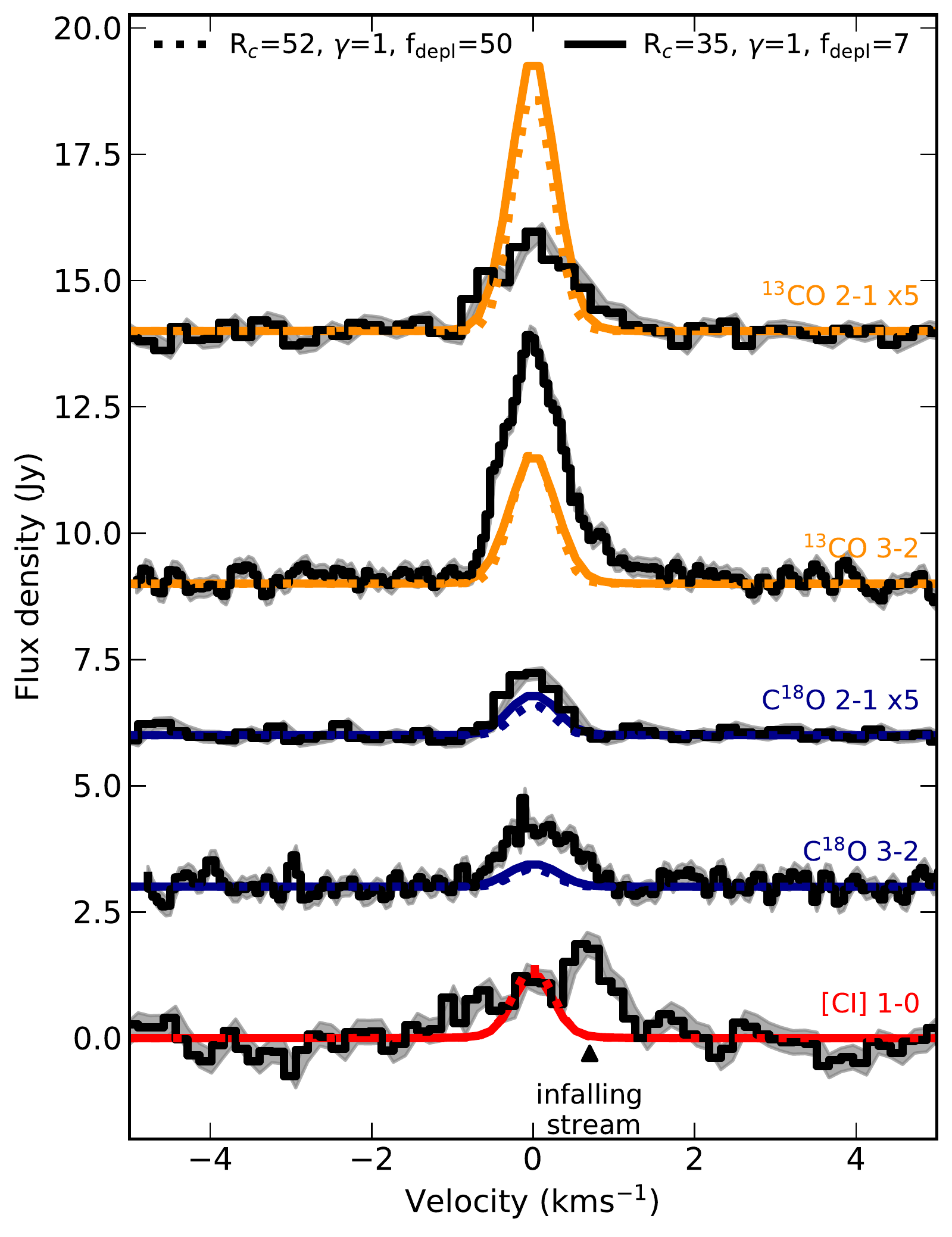}
    \caption{Line profiles for DR Tau, same as Fig. \ref{fig:do_lines}, but with changes in $\gamma=1$ and $R_\mathrm{c}$ = 35, respectively, as indicated in the Figure legend.}
    \label{fig:dr_lines_smallrc}
\end{figure}

DR Tau is a very complex source with a lot of substructure in the different gas tracers.
We explored a range of modeling parameters that fit the SED reasonably well, but all of the models have their own caveats.
Here we present some additional models to give a feeling for the uncertainty in carbon abundance of this source-specific model.
Unfortunately, there is no archival high resolution $^{12}$CO data available and the isotopologues $^{13}$CO and C$^{18}$O J = 2-1 are too weak to determine the spatial scale of the gas disk, as the outer regions of the disk are not detected with high confidence. 
This means that the spatial information we have of DR Tau is limited.
In Fig. \ref{fig:dr_profile} A. we present the azimuthally averaged radial profile of the normalized flux of the \cott $J$ = 3-2 emission, together with a small grid of models.
In Fig. \ref{fig:dr_lines_smallrc} we present the corresponding emission spectra of the different transitions.
We lowered the carbon abundance to match the modeled [C~I] line flux with our observations.
Our fiducial model has $R_\mathrm{c}$ = 52 AU, similar to the continuum observations, and $\gamma$ = 1.
This model fits the spatial extent of the \cott emission (Fig. \ref{fig:dr_profile} A) reasonably well, but under-predicts the line fluxes of most CO isotopologue transitions (Fig. \ref{fig:dr_lines_smallrc}). 
Using this model we find volatile outer disks carbon depletion fractions up to 50 to match the [C~I] $J$ = 1-0 observations, while a depletion fraction of 5 suffices to fit the CO isotopologue transitions.


Following the procedure described in \citet{2018A&A...619A.113M}, we fitted a powerlaw surface density profile in the so called pivot-region of the radial profile of the CO isotopologue $J$ = 3-2 emission, where the line is optically thin, but has enough column density that isotope selective processes are not a dominant factor. 
We present the results of this fit in Fig. \ref{fig:dr_profile} B.
For \cott we performed a linear least-squares fit between 75 - 175 AU.
For \coet the pivot region extends from 30 - 150 AU.
We find $\gamma$ = 0.5 and $\gamma$ = 0.7 for \cott and \coet, respectively, strengthening the case that $\gamma$ < 1 for DR Tau.

The best representative model of the line fluxes, presented in the main text, has $\gamma$ = 0.5 and $R_\mathrm{c}$ = 52 AU, with the same mass and geometry as the fiducial model.
This model has a smaller spatial extent than the \cott observations (Fig. \ref{fig:dr_profile} A), but fits both the CO isotopologue data as well as the [C~I] line fluxes (Fig. \ref{fig:dr_lines}).
We tested a model with $\gamma = 1$ and a smaller disk, $R_\mathrm{c}$ = 35 AU, that matches the spatial distribution of the CO isotopologue emission of the model with $\gamma$ = 0.5 and $R_\mathrm{c}$ = 52 AU, but this model underproduces the CO isotopologue emission with respect to the [C~I] emission, similar as the fiducial model.
The fit could be improved in the future by using a different sized gas disk than the dust disk, as follows from observations, which is beyond the scope of this paper. 

In all possible models that we analyzed the \cott $J$ = 2-1 flux is over-predicted by a factor of a few, which is probably due to the low UV-sampling at shorter baselines of the data.
We neglect this line further in our modeling.
In the main text we used the model that is closest to the continuum observations and reproduced the line emission spectra best. The effect of under-predicting the size of the disk, if any, could be that the disk is more depleted in volatile carbon than assumed in the text, but less than $f_\mathrm{depl}$=50.

\section{Channel maps}
\begin{figure*}
\includegraphics[width = 1\linewidth]{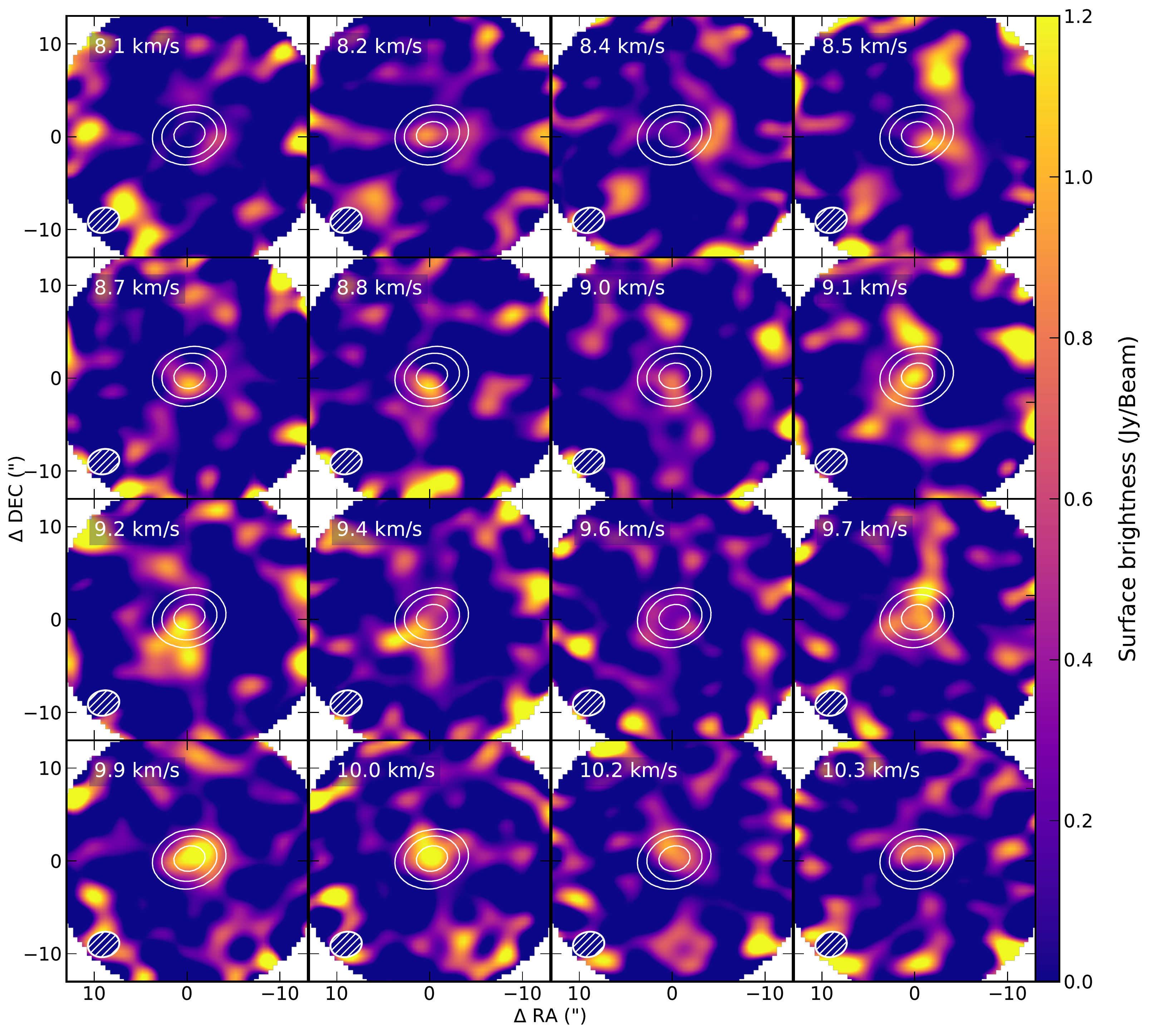}
\caption{Primary beam corrected channel maps of the [C~I] emission in DR Tau. Solid white contours show the millimeter continuum at 10,50 and 200 $\sigma$, respectively.}
\label{fig:dltaucube}
\end{figure*}

\begin{figure*}
\includegraphics[width = 1\linewidth]{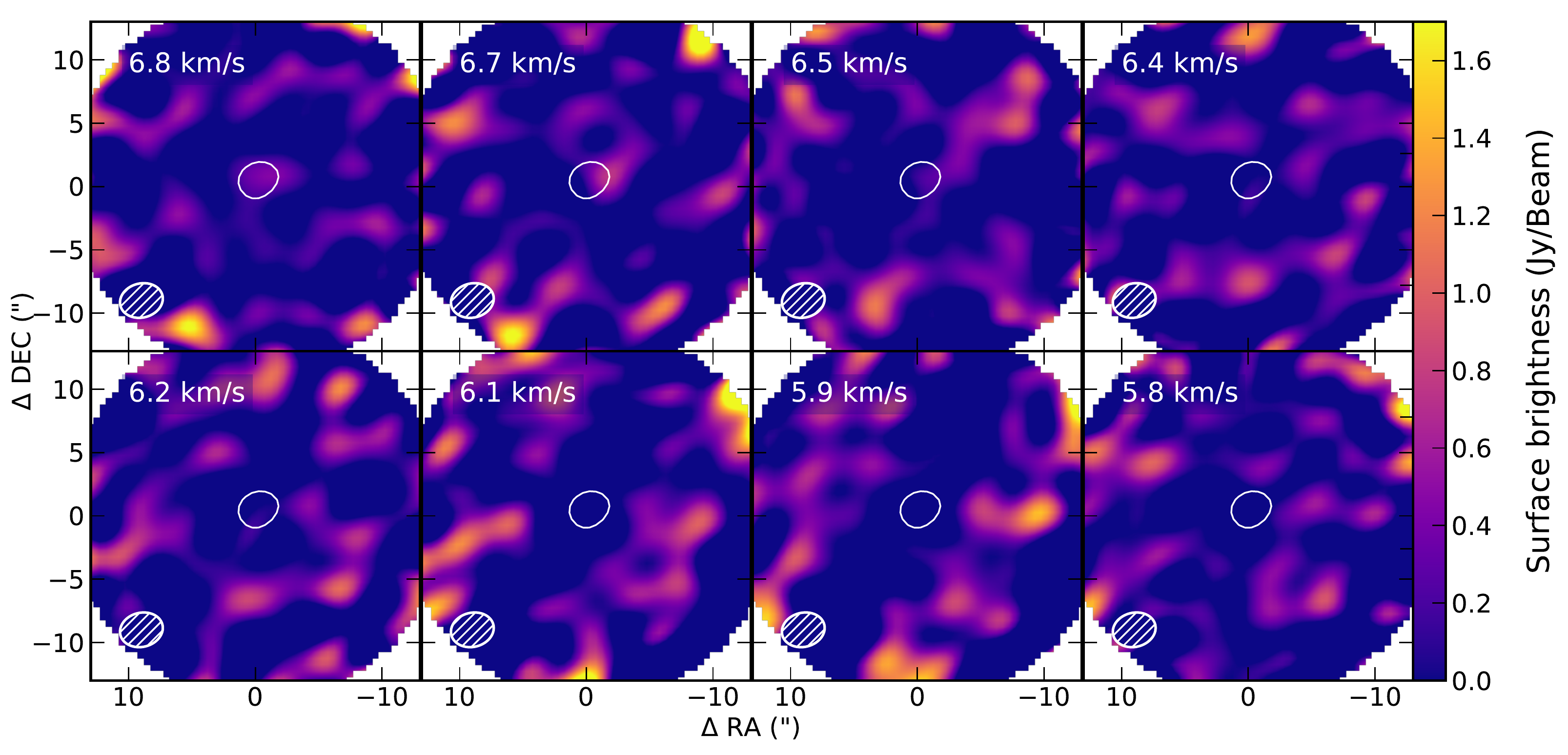}
\caption{Same as Fig. \ref{fig:dltaucube} for FZ Tau.}
\label{fig:fztaucube}
\end{figure*}

\begin{figure*}
\includegraphics[width = 1\linewidth]{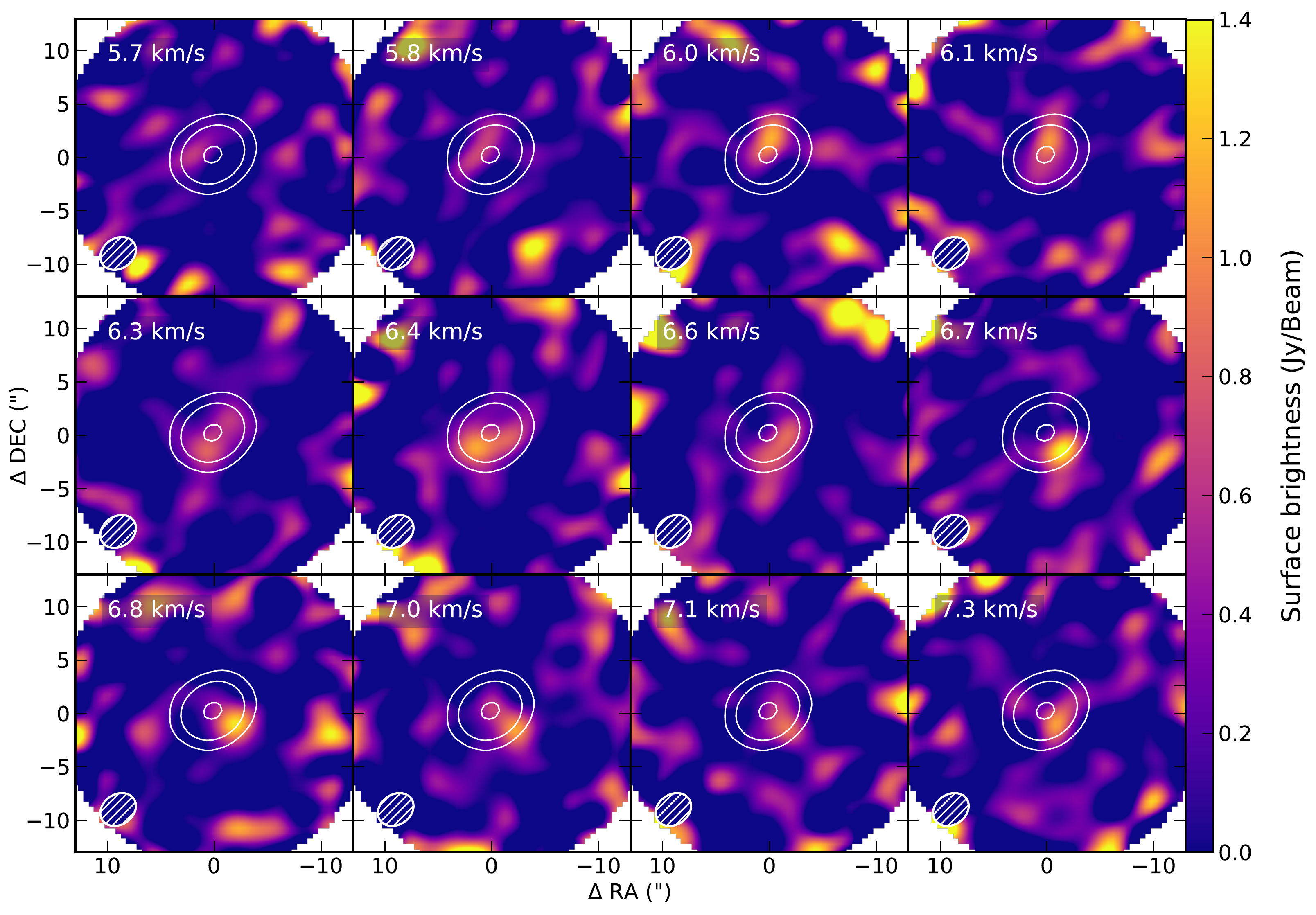}
\caption{Same as Fig. \ref{fig:dltaucube} for DL Tau.}
\label{fig:drtaucube}
\end{figure*}

\begin{figure*}
\includegraphics[width = 1\linewidth]{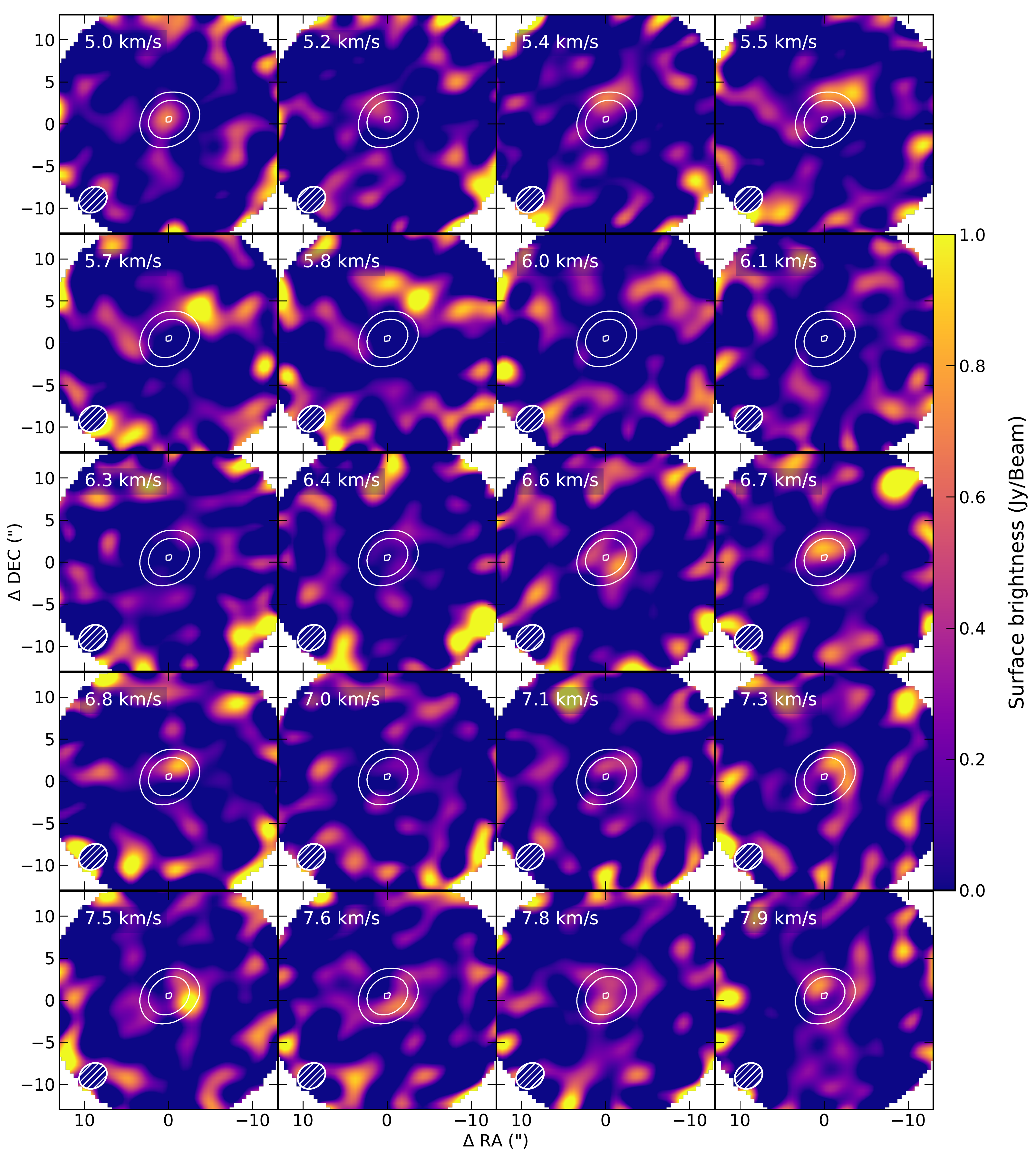}
\caption{Same as Fig. \ref{fig:dltaucube} for DO Tau.}
\label{fig:dotaucube}
\end{figure*}

\begin{figure*}
\includegraphics[width = 1\linewidth]{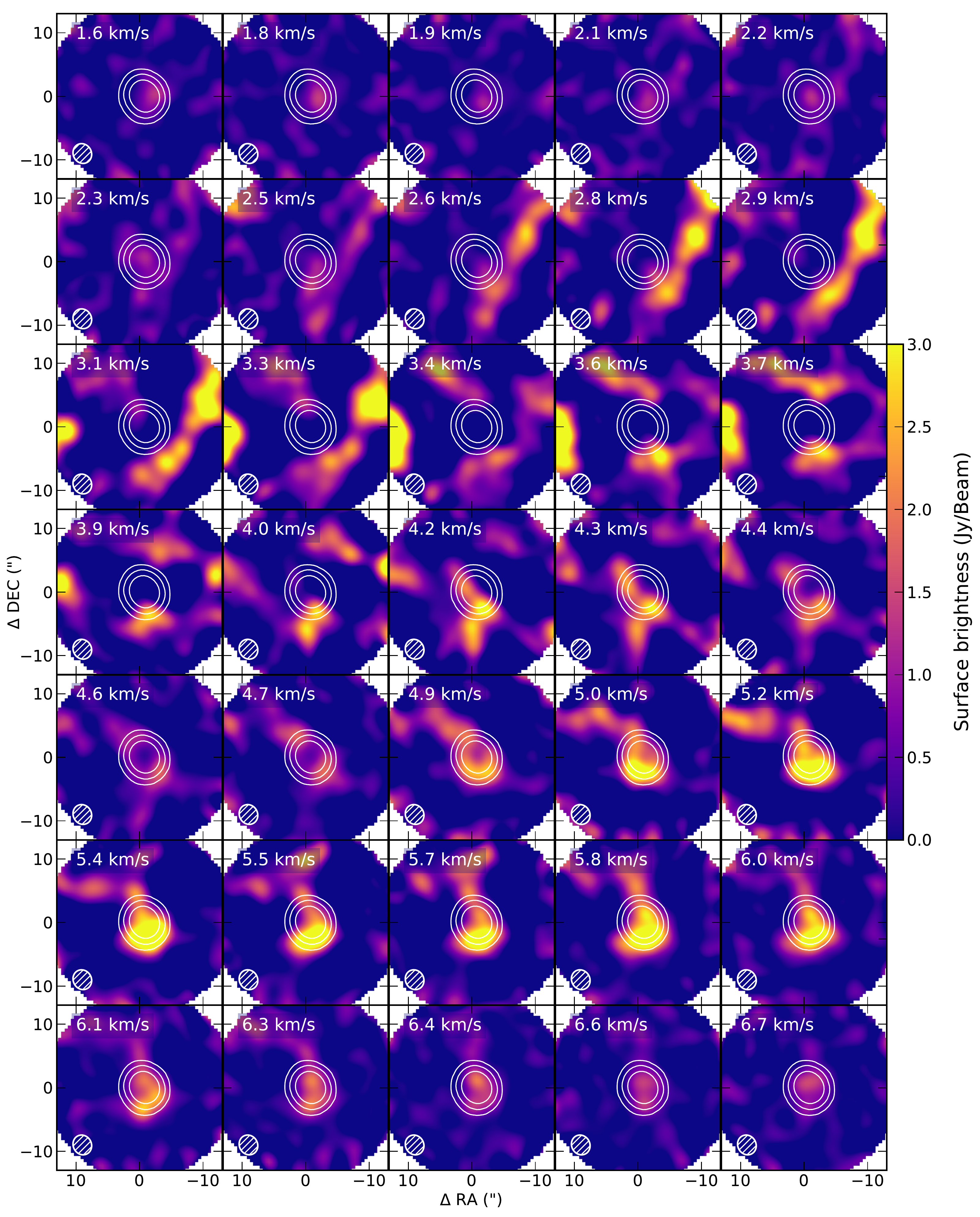}
\caption{Same as Fig. \ref{fig:dltaucube} for WW Cha}
\label{fig:wwchacube}
\end{figure*}

\begin{figure*}
\includegraphics[width = 1\linewidth]{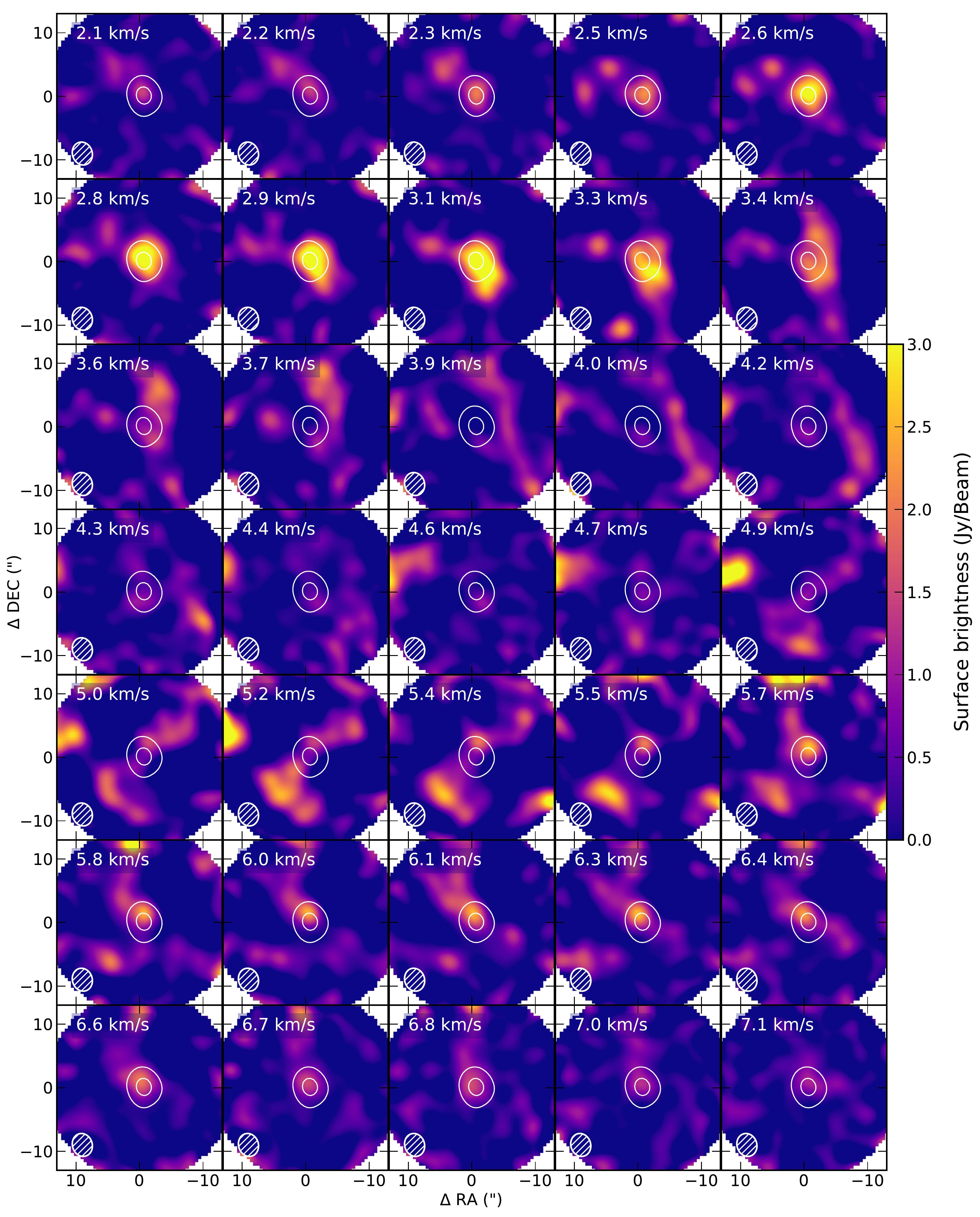}
\caption{Same as Fig. \ref{fig:dltaucube} for FM Cha.}
\label{fig:fmchacube}
\end{figure*}

\begin{figure*}
\includegraphics[width = 1\linewidth]{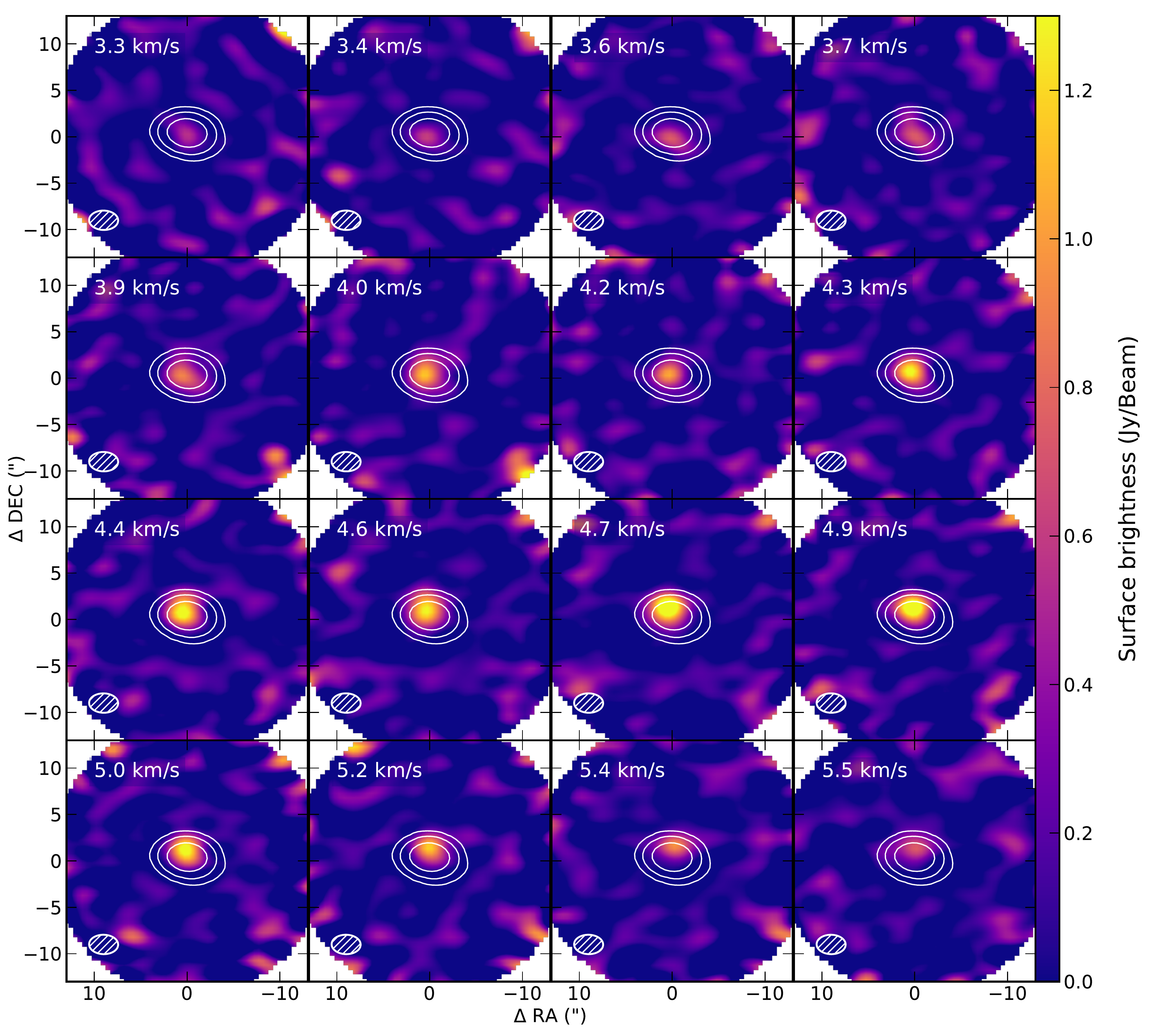}
\caption{Same as Fig. \ref{fig:dltaucube} for AS 205 A.}
\label{fig:as205cube}
\end{figure*}

In Fig. \ref{fig:dltaucube} - Fig. \ref{fig:as205cube} we present the channel maps of the [C~I] observations for each source.

\end{appendix}
\end{document}